\documentclass{llncs}
\usepackage[utf8]{inputenc}
\usepackage[english]{babel}
\usepackage{amsmath}
\usepackage{amssymb}  
\usepackage{stmaryrd} 
\usepackage{latexsym} 
\usepackage{xcolor}
\usepackage{import}
\usepackage{array}
\usepackage{multirow}
\usepackage{tikz}

\usepackage{listings}

\lstdefinestyle{code}{
  basicstyle=\fontfamily{txtt}\fontseries{m}\small\selectfont\color{blue},
  lineskip=-1pt,
}

\lstdefinestyle{bluecode}{
  basicstyle=\fontfamily{txtt}\fontseries{m}\small\selectfont\color{blue},
  lineskip=-1pt,
  keywordstyle=\color{blue}
}

\usepackage[backref,pagebackref=true]{hyperref}
\hypersetup{%
  pdftitle={The Unfolding Semantics of Functional Programs},%
  pdfauthor={Jos{\'e}-Mar{\'i}a Rey and Julio Mari{\~n}o},%
  bookmarksopen=true,
  colorlinks=true,%
  linkcolor=red,%
  citecolor=blue,
  anchorcolor = red,
  pagecolor=red,
  urlcolor=blue
}

\begin{document}
\pagestyle{headings}  
\title{The Unfolding Semantics of Functional Programs}
\titlerunning{Unfolding Semantics}  
%
\author{Jos{\'e} M.~{Rey Poza}
  \and Julio Mari{\~n}o
}
\authorrunning{Rey, Mari{\~n}o} 
\institute{School of Computer Science,\\
           Universidad Politécnica de Madrid\\
\email{josem.rey@gmail.com, jmarino@fi.upm.es}
}

\maketitle              

\newcommand{\TC}{\mathit{TC}} 
\newcommand{\DC}{\mathit{DC}} 
\newcommand{\FS}{\mathit{FS}} 
\newcommand{\CP}{\mathit{CP}} 
\newcommand{\CF}{\mathit{CF}} 
\newcommand{\PF}{\mathit{PF}} 
\newcommand{\Vars}{\mathit{Vars}} 
\newcommand{\CE}{\mathit{CE}} 
\newcommand{\CS}{\mathit{CS}} 
\newcommand{\LC}{\mathit{LC}} 
\newcommand{\AC}{\mathit{AC}} 
\newcommand{\Herb}{${\cal H}$}
\newcommand{\Set}{\ensuremath{{\cal S}\!\mathit{et}}}
\newcommand{\Mypfun}{\ensuremath{\rightarrow\!\!\!\! \shortmid~~}}
\newcommand{\sbot}{\ensuremath{\bot^s}} 
\newcommand{\MyIfThenOp}{\ensuremath{\blacktriangleright}}
\newcommand{\MyIfThen}{\ensuremath{\triangleright}}

\begin{abstract}
The idea of using unfolding as a way of computing a program semantics
has been applied successfully to logic programs and has
shown itself a powerful tool that provides concrete, 
implementable results, as its outcome is actually source code.
Thus, it can be used for
characterizing \emph{not-so-declarative} constructs in \emph{mostly}
declarative languages, or for static analysis. However,
unfolding-based semantics has not yet been applied to higher-order,
lazy functional programs, perhaps because some functional features 
absent in logic programs make the correspondence between
execution and unfolding not as straightforward. 
This work presents an unfolding semantics for higher-order, lazy
functional programs and proves its adequacy with respect to a given
operational semantics. Finally, we introduce some applications of our
semantics. 
\end{abstract}
\section{Introduction}

The broad field of program semantics can be classified according to
the different \emph{meanings} intended to be captured or the various
techniques employed. Thus, traditionally, the term \emph{denotational
  semantics} is used when a high-level, implementation
independent description of the behaviour of a program is pursued, while
\emph{operational semantics} usually refers to descriptions intended
to capture more implementation-related properties of the execution of
a program, which can then be used to gather resource-aware
information, or as a blueprint for actual language implementations.%


The inability of those denotational semantics to capture certain
aspects of logic programs (such as the \emph{computed answer}
semantics) and the introduction of ``impure'' constructs in Prolog,
led to a considerable amount of proposals for alternative semantics of
logic programs during the 80's and 90's. One of the most remarkable
proposals is the so-called \emph{s-semantics}
approach~\cite{Bossi94s-semanticsapproach} which explores the
possibility of using \emph{syntactic denotations} for logic
programs. In other words, programs in a very restricted form are the
building blocks of the denotation, and program transformation (e.g.~via
\emph{unfolding}) takes the role of interpretation transformers in
traditional constructions. Being closer to the source code facilitates
the treatment of the less declarative aspects. 

However, in spite of the fact that unfolding is a technique equally
applicable to functional programs, little attention has been paid to
its use as a semantics tool. Investigating how unfolding can be
applied to find the semantics of functional programs is the goal of this
paper.

\subsection{Unfolding Semantics}
\label{subsec_unfolding}
The process of unfolding is conceptually simple: replace any function
or predicate invocation by its definition. In logic programming this
amounts to unifying some literal in the body of a rule with the head
of some \emph{piece of knowledge} that has already been calculated,
and placing the corresponding body instance where the
literal was.

The previous paragraph mentions two important concepts: the first is
that of {\em   piece of knowledge} generated by unfolding program
rules according to all current pieces of knowledge. Every piece of
knowledge (called a {\em fact}) is valid source code. The set of facts
may increase with every iteration. A set of facts is called an {\em
  interpretation}. In addition, the second concept hinted in the
paragraph above is that of initial interpretation. 

\begin{figure}[t]
\begin{center}
\begin{tabular}{m{6.25cm} | m{6cm}}
{\bf Prolog code / \emph{s}-semantics unfolding} & {\bf Functional
  code / Funct.~unfolding} \\
\lstset{style=bluecode}
\begin{lstlisting}
add(zero,X,X).
add(suc(X),Y,suc(Z)):-add(X,Y,Z).
\end{lstlisting}

&

\begin{lstlisting}
add Zero x = x
add (Suc x) y = Suc (add x y)
\end{lstlisting}
\lstset{style=code}
\\

\begin{tabular}{p{0.75cm} p{4.75cm}}
$S_1=$ & 
\{\texttt{add(zero,X,X)}\} \\
$S_2=$ &
\{\texttt{add(zero,X,X)}, \\ 
 &  \texttt{add(suc(zero),X$_2$,suc(X$_2$))}\} \\
\end{tabular}

&

\begin{tabular}{p{0.75cm} p{4.75cm}}
$I_0=$ & $\emptyset$ \\
$I_1=$ & 
\{\texttt{add~Zero~x~=~x}\} \\
$I_2=$ & 
\{\texttt{add~Zero~x~=~x}, \\
  &    \texttt{add~(Suc~Zero)~y = (Suc~y)}
\} \\
\end{tabular}

\end{tabular}
\end{center}
\caption{Logic and functional versions of a simple program, and their
  unfoldings.} 
\label{fig_original}
\end{figure}

\subsubsection{Unfolding in Logic Programming.}
As an example, the left half of Fig.~\ref{fig_original}  
shows a predicate called \lstinline.add. that adds two Peano
Naturals. This part shows the source code (upper
side) together with the corresponding unfolding results (lower side).

The general unfolding procedure can be easily followed in the example,
where the first two clause sets are generated ($S_1$ and $S_2$).

\subsubsection{Unfolding in Functional Programming.}
Unfolding in functional programming (FP) follows the very same idea
of unfolding in logic programming: any function invocation is replaced
by the right side of any rule whose head {\em  matches} the invocation.

Consider the right half of Fig.~\ref{fig_original} as the functional
version of the previous example, written in our model language. Some
differences and analogies between both paradigms can be spotted: 
In FP, unfolding generates rules (equations)
as pieces of knowledge, instead of clauses which appeared in logic programming.
The starting seed is also different: bodyless rules are used in logic
programming while the empty set is used in functional programming.

Finally, observe that both unfoldings (logic and functional) produce
valid code and analogous results, being $I_n$ {\em equivalent} to
$S_n$. This fact provides a clue into two of the main reasons to
define an unfolding semantics: first they are {\em implementable} as
the procedure above shows and, second, they are also a clear point between
denotational and operational semantics in proving the equivalence
between a semantics of each type.

\subsection{Extending Unfolding Semantics to Suit Functional Programs}

\label{subsec_nottrivial}
\begin{figure}[t]
\begin{tabular}{m{5cm} m{8cm}}
{\bf Functional code} & {\bf Unfolding} \\
\lstset{style=bluecode}
\begin{lstlisting}
ite : Bool -> a -> a -> a
ite True  t e = t
ite False t e = e

filter:(a->Bool)->[a]->[a]
filter p [] = []
filter p (x:xs) =
  ite (p x) 
      (x:(filter p xs))
      (filter p xs) 
\end{lstlisting}
\lstset{style=code}
&
  \begin{tabular}{p{0.2cm} p{0.2cm} p{8.5cm}}
  $I_0$ & = & $\emptyset$ \\
  $I_1$ & = & \{\texttt{ite(True,t,e) = t}, 
                \texttt{ite(False,t,e) = e}, \\
        & &     \texttt{filter(b,Nil) = Nil}\} \\
  $I_2$ & = & \{\texttt{ite(True,t,e) = t}, 
                \texttt{ite(False,t,e) = e}, \\
        & &     \texttt{filter(b,Nil) = Nil}, \\
        & &     \texttt{filter(b,Cons(c,Nil))|} \\
        & &     \texttt{  snd(match(True,b@[c]))=Cons(c,Nil)}, \\
        & &     \texttt{filter(b,Cons(c,Cons(d,e)))|} \\
        & &     \texttt{  snd(match(True,b@[c]))=Cons(c,Bot)}, \\
        & &     \texttt{filter(b,Cons(c,Nil))|} \\
        & &     \texttt{  snd(match(False,b@[c])) = Nil} \\
        & &     \texttt{filter(b,Cons(c,Cons(d,e)))|} \\
        & &     \texttt{  snd(match(False,b@[c]))=Bot}\} 
  \end{tabular}
\end{tabular}
\caption{Functional program requiring higher-order applications.}
\label{fig_higher}
\end{figure}

Section \ref{subsec_unfolding} showed that the ideas of unfolding
semantics in logic programming can also be applied 
to FP. However, some features of FP (e.g.~higher-order, laziness) render the
 unmodified procedures invalid.   

Consider the function \lstinline.filter. in Fig.~\ref{fig_higher}. It
takes a list of values and returns those values in the list that
satisfy a predicate passed as its first argument.

Applying naïve unfolding to \lstinline.filter. is impossible since
\lstinline.ite. (short for if-then-else) demands a 
boolean value but both \lstinline.p.  
and \lstinline.x. are unknown at unfold time (i.e.~before execution). 

In order to overcome this limitation, we have developed a technique
capable of generating facts in the presence of incomplete information. 
In this case we generate \emph{conditional} facts 
(the last four facts in $I_2$). 
The function \lstinline.match. checks whether a given term {\em matches} an
expression that cannot be evaluated at unfolding time (here, 
\lstinline.(p x)).. Observe that \lstinline.match. must be ready to
deal with infinite values in its second argument.

Note that, in automatically-generated code, such as the unfolded code
shown in Fig.~\ref{fig_higher} and the figures to come, variables are
most often renamed and that our unfolding implementation uses tuples
to represent curried expressions.  

\subsection{Related Work}
\label{subsec_related}

One of the earliest usages of unfolding in connection to semantics is
due to Scott \cite{Scott70}, who used it to find the denotation of
recursive functions, even though the word {\em unfolding} was not used
at the time.

Concerning logic programming, our main inspiration source is \emph{s}-semantics
\cite{Bossi94s-semanticsapproach}, which defines an unfolding
semantics for pure logic programs that is defined as the set of literals
 that can be successfully derived by using the program given.

In addition, fold and unfold have been used in connection to many
other problems in the field of logic programming. For example
\cite{Pettorossi00perfectmodel} describes a method to check whether 
a given logic program verifies a logic formula. It does this by
applying program transformations that include fold and unfold.

Partial evaluation of logic programs has also been tackled by means of
unfolding but it usually generates huge data structures even for
simple programs. 

As in logic programming, fold/unfold transformations have been used
extensively to improve the efficiency of functional
programs~\cite{Burstall:1977:TSD:321992.321996}, but not 
as a way of constructing a program's semantics.
 

Unfolding has also been applied to functional-logic programming
\cite{Alpuente96narrowing-drivenpartial}. However, that paper is not
oriented towards finding the meaning of a program but to 
unfold it partially to achieve some degree of partial evaluation.
Besides, it is restricted to first order, eager languages.

\paragraph{Paper Organization} 
Section 2 presents preliminary concepts. Section 3 describes the unfolding
semantics itself, the core of our proposal. Section 4 presents the
formal meaning we want to assign to the core language that we will be
using. Section 5 points out some applications of unfolding
semantics. Section 6 concludes. 

\section{Preliminaries}

\paragraph{Notation}
\label{subsec_notation}
Substitutions will be denoted by $\sigma, \rho, \mu$. $\sigma(e)$ or
just $\sigma e$ will denote the application of 
substitution $\sigma$ to $e$. The empty substitution will be denoted
by $\epsilon$. $e \equiv e'$ will denote that
the expressions $e$ and $e'$ have the same syntax tree. 

Given a term $t$, a position within $t$ is denoted by a dot-separated
list of integers. $t|_o$ denotes the content of position $o$ within
$t$. Replacement of the content at position $o$ within a term $t$ by some
term $t'$ is denoted by $t[t']|_o$. The set of positions within an expression
$e$ will be denoted by $\mathit{Pos(e)}$.

$k, d$ will be used to denote constructors while $c$ will denote guards. 

The {\em auxiliary} functions $\mathit{fst}:a \times b \rightarrow a$ and 
$\mathit{snd} : a \times b \rightarrow b$ extract the first and second
element of a tuple, respectively. Boolean conjunction and disjunction
are denoted by $\wedge$ and $\vee$. $\mathit{mgu}(f~t_1 \ldots
t_n,f~e_1 \ldots e_n)$ (where the $t_j$ are terms and $e_i$ do not
have user-defined functions) denotes its most general unifier.
The conditional
operator will denoted by $\MyIfThenOp$, which has type $(\MyIfThenOp) :
\mathit{Bool} \rightarrow a \rightarrow a$ and is defined as:
$(\mathit{True} \MyIfThenOp a) = a$, $(\mathit{False} \MyIfThenOp a) =
\sbot$. 

Regarding types, $A \Mypfun B$ denotes a partial function from domain
$A$ to domain $B$. The type of $\pi$-interpretations (sets of
facts) is noted by $\Set({\cal F})$. ${\cal F}$ is intended to denote the domain
from which facts are drawn. The projection of an interpretation $I$
to predefined functions only is denoted as $I_p$. 
Lack of information is represented by $\sbot$ in unfolding
interpretations and by the well-known symbol $\bot$ when it is related
to the minimum element of a Scott domain. 
%
Lastly, HNF stands for \emph{head normal form}. An expression is said to be
in head normal form if it is a variable or its root symbol is a constructor.
\emph{Normal form} (NF) terms are those in HNF, and whose subterms are
in NF.

\subsection{Core Language. Abstract Program Syntax}
\label{subsec_abstract_program_syntax}

The language\footnote{We assume the core language to
be typed although we do not develop its typing discipline here
because of lack of space.}  
that will be the base to develop this work is a functional language
with guarded rules. Guards (which are optional) are boolean
expressions that must be rewritten to \lstinline.True. in order for
the rule that contains it to be applicable. 

Note that the language we are using is a purely functional language
(meaning that it uses pattern matching,higher-order features and referential
transparency).   


Let us consider a signature $\Sigma=\langle V_\Sigma, \DC_\Sigma,
\FS_\Sigma, \PF_\Sigma \rangle$ where $V$ is the set of variables,
$\DC$ is the set of Data Constructors that holds at least $\sbot$ and
a tuple-building constructor, $\FS$ holds the user-defined functions
and $\PF$ denotes the set of predefined functions that holds at least
 a function \lstinline.match., a function {\em nunif} and a function
 \lstinline.@. that applies an expression $e$ to a list of expressions (that is,
$e$\lstinline.@.[$e_1,\ldots,e_n$] represents $(e~e_1 \ldots
e_n)$). $\PF$ and $\FS$ are disjoint. 

Some of the sets above depend on the program
  $P$ under study, so they should be denoted as, e.g., $\FS_P$ but we will
  omit that subscript if it is clear from the context.
All these sets are {\em indexed} by arity. The
domains for a program are:
\[
\begin{array}{l@{~}rcl@{~~~}l}
\textsc{(Variables)}   & V & ::= & x, y, z, w\dots \\
\textsc{(Terms)}  & T & ::= & v ~|~ k~t_1 \dots t_k & v \in V,  
k \in \DC, t_i \in T\\ 
\textsc{(Expressions)} & E & ::= & t ~|~ f'~e_1 \dots e_k  & f' \in \FS \cup \PF, 
t \in T, e_i \in E \\ 
\textsc{(Patterns)}    & \mathit{Pat} & ::= & f~t_1 \dots t_k & f \in
\FS, t_i \in T \\
\textsc{(Rules)}       & \mathit{Rule} & ::= & l ~\verb.|.~ g = r & l \in
\mathit{Pat}, g \in E, r \in E \\
\textsc{(Programs)}    & {\cal P} & ::= & \Set(\mathit{Rule})
\end{array}
\]

\noindent
Terms are built with variables and constructors only. Expressions
comprise terms and those constructs that
include function symbols (predefined or not).

Note that the description corresponding to expressions $(E)$ does not allow for
an expression to be applied to another expression but we still want
our language to be a higher order one. We manage higher order by means
of partial applications, written by using the predefined function
\lstinline.@.. Thus, un application like $e_1 e_2$ (where $e_1$ and
$e_2$ are arbitrary expressions) is represented in our setting by
$e_1$\lstinline.@.$[e_1]$ (or by \lstinline.@.($e_1$,$[e_2]$) in
prefix form). 

To ensure that programs defined this way constitute a
confluent rewriting system, these restrictions will be imposed
on rules \cite{Hanus94theintegration}: linear rule patterns,
no free variables in rules (a free variable is
one that appears in the rule body but not in the guard or the pattern)
and finally, no superposition among rules 
(i.e. given a function application, at most a rule must be applicable). 

The core language does not include local declarations (i.e. {\em let,
  where}) but this does not remove any power from the language since
local declarations can be translated into aditional rules by means of
lambda lifting.

\section{Unfolding Semantics for the Core Language}
\label{sec_unfolding_sem}

\subsection{Interpretations}

\begin{definition}[Fact and $\pi$-Interpretation]
\label{def_fact_interpretation}
We will use the word {\em fact} to denote any piece of
  {\em proven knowledge} that can be extracted from a program $P$ and
  which conforms to  
the following restrictions: (i) They have shape $h ~|~ c = b$, (ii)
$b$ and $c$ include no symbols 
belonging to $\FS$, (iii) Predefined functions are not allowed inside
$b$ or $c$ unless 
the subexpression headed by a symbol in $\PF$ cannot be evaluated further 
(e.g.~$x+1$ would be allowed in $b$ or $c$ but $1+1$ would not, $2$ should be
used instead) and (iv) The value of $c$ can be made equal to 
 \lstinline.True. (by giving appropriate values to its variables).
The type of facts is denoted ${\cal F}$. Facts can be seen as rules
with a restricted shape.

In addition, a $\pi$-interpretation is any set of valid facts that can
be generated by using the signature of a given program $P$. The concept of
$\pi$-interpretation has been adapted from the concept with the same
name in \emph{s}-semantics. 
\end{definition}

The reason for imposing these restrictions on facts is to have some
kind of {\em canonical form} for interpretations. Even with this
restrictions, a program does not have a unique interpretation, but we
intend to be as close to a canonical form for interpretations as possible.

\subsection{Defining the Unfolding Operator}

The process we are about to describe is represented in pictorial form
in the Appendix, Sect.~\ref{sec_unfolding_pictorial} in order to help
understand the process as a whole.

The unfolding operator relies on a number of auxiliary functions that
are described next, together with the operator itself. A full example
aimed at clarifying how these functions work can be found in the
Appendix (Example \ref{ex_global}).  

\subsubsection{Evaluation of Predefined Functions}
\label{sec_eval}

\begin{figure}
  \begin{center}
    $\mathit{eval} : \Set({\cal F}) \times E \rightarrow E$
  \end{center}

  \begin{tabular}{m{3.1cm}m{2mm}m{3cm}l} 
  $\mathit{eval}(I,x)$ & = & $x$ & $x \in V$
  \\
  $\mathit{eval}(I,(k~e_1 \ldots e_n))$ &=&  
  $(k~e'_1 \ldots e'_n)$ & $(k \in \DC, n \geq 0,$
  $\mathit{arity}(k)=n, \mathit{eval}(I,e_i)=e'_i)$ 
  \\
  $\mathit{eval}(I,(k~e_1,\ldots,e_n))$ & = &
  $ (k~e_1 \ldots e_n)$ & $(k \in \DC, n \geq 0,
  n<\mathit{arity}(k))$ 
  \\
  $eval(I,(p~e_1 \ldots e_n))$ & = & $(I_p~e'_1 \ldots e'_n)$ 
  &
  if $(p~e'_1,\ldots,e'_n)$ can be evaluated to NF \\ 
  & & & without error. It is left untouched otherwise. \\
  & & & $(\mathit{eval}(I,e_i)=e'_i, $    
  $p \in \PF - \{\mathit{match}\})$. \\
  $\mathit{eval}(I,(c_1 \wedge \ldots \wedge c_{i-1} \wedge
  \mathit{snd}(\mathit{match}(p,e)) \wedge$
   & = & & $\sigma(c_1 \wedge \ldots \wedge c_{i-1} 
    \wedge b \wedge c_{i+1} \wedge \ldots \wedge
    c_n \MyIfThenOp e')$
  \\
    $c_{i+1} \wedge \ldots \wedge c_n \MyIfThenOp e'))$ & & &
     if $\mathit{match}(p,e)=(\sigma,b)$. \\
  $\mathit{eval}(I,(x~e_1 \ldots e_m))$ & = & $(x~e_1 \ldots e_m)$ &
  $(x \in V)$ 
  \\
  $\mathit{eval}(I,(p~e_1 \ldots e_m))$ & = & $
  (p~e_1 \ldots e_m)$ & $p \in \PF_o, m<o$. 
  \\
  $\mathit{eval}(I,(f~e_1 \ldots e_m))$ & = & $(f~e_1 \ldots e_m)$ & $f
  \in \FS_n, m \leq n$ 
  \\
  \end{tabular} 
  \caption{Evaluation of predefined functions}
  \label{fig_eval}
\end{figure}

The function {\em eval} (Fig.~\ref{fig_eval}) is in charge
of finding a value for those expressions which do not contain any full
application of user-defined functions. Since predefined functions do
not have rules, their appearances cannot be rewritten, just
evaluated. Only predefined functions are evaluated; all the other
expressions are left untouched. Note that $\mathit{eval}$ requires the
interpretation in order to know how to evaluate predefined functions. 

\subsubsection{Housekeeping the Fact Set}
\label{sec_clean}

Every time a step of unfolding takes place, new facts might be added
to the interpretation. These new facts may overlap with some existing
facts (that is, be applicable to the same 
expressions as the existing ones). Although overlapping facts do not
alter the meaning of the program, they are redundant and removing them
makes the interpretation smaller and more efficient. The function {\em
  clean} removes those redundancies. We believe this {\em cleaning
  step} is a novel contribution in the field of unfolding semantics
for functional languages (see \cite{Alpuente97}, where a
semantics for logic functional programs is presented but where {\em
  interpretations} are not treated to remove any possible overlapping).

\noindent
Given an interpretation, the function $\mathit{clean}$ removes
the overlapping pairs in order to erase
redundant facts. Before defining {\em clean}, some definitions are needed.

\begin{definition}[Overlapping Facts]
\label{def_overlapping_facts}
A fact $h ~|~c = b$ overlaps with some other fact $h'
~|~ c' = b'$ if the following two conditions are met:  

\begin{itemize}
\item
There exists a substitution $\mu$ such that: $h \equiv \mu(h')$ and

\item
The condition $c \wedge \mu(c')$ is satisfiable\footnote{Note that
  satisfiability is undecidable in general. This means that there
  might be cases where clean is unable to remove overlapping facts.}.
\end{itemize}
\end{definition}

\noindent
Intuitively, two facts overlap if there is some expression that can be
rewritten by using any of the facts. 

What {\em clean} does is to remove any overlapping between facts from
the interpretation it receives as argument. It does this by conserving
the {\em most specific fact} of every overlapping fact set untouched
while restricting the other facts of the set so that the facts do not
overlap any more. This restriction is accomplished by adding new
conditions to the fact's guard.   

In order to be able to state that a fact is more specific than some
other, we need an ordering:  

\begin{definition}[Term and Fact Ordering]
\label{def_term_fact_order}
Let us define $t \sqsubseteq t' ~(t,t' \in E), \\  
(t,t') \mathit{~linear~or~} \mu(t)=\sigma(t')
\mathit{~for~some~substitutions~} \mu, \sigma$:
\begin{itemize}
\item
$\sbot \sqsubseteq t ~~ t \in T$

\item
$t \sqsubseteq x ~~t \in T, x \in V$

\item
$t_1 \ldots t_n \sqsubseteq t'_1 \ldots t'_n$ if and only if $t_i \sqsubseteq t'_i ~~ \forall i:1..n$

\item
$(k~t_1 \ldots t_n) \sqsubseteq (k~t'_1 \ldots t'_n)$ if and only if $t_i \sqsubseteq t'_i ~~ \forall i:1..n, k \in \DC \cup \PF$
\end{itemize}

\noindent
Now, this ordering can be used to compare facts. 

Given two overlapping facts 
$F \equiv f~t_1 \ldots t_n ~|~c = b$ and $F' \equiv
f~t'_1 \ldots t'_n ~|~c' = b'$, it is said that $F'$ is more specific
than $F$ if and only if at least one of the following criteria is met: 

\begin{itemize}
\item
$t'_1 \ldots t'_n \sqsubset t_1 \ldots t_n$ or

\item
If $t'_1 \ldots t'_n$ and $t_1 \ldots t_n$ are a variant of each
other (i.e., they are the same term with variables renamed), the fact
that is more specific than the other is the one with the most 
restrictive guard (a guard $c'$ is more restrictive than 
another guard $c$ if and only if $c'$ entails $c$ but not
viceversa).

\item
If two facts are such that their patterns are a variant of each
other and their guards entail each other, the fact that is more
specific than the other is the
one with the greatest body according to $\sqsubseteq$. 
\end{itemize}
\end{definition}

\noindent
Remember that facts' bodies do not contain full applications of user-defined
functions, so $\sqsubseteq$ will never be used to compare full
expressions. However, $\sqsubseteq$ may be used to compare expressions
with partical applications or with predefined functions. In these
cases, function symbols (both from $\FS$ or from $\PF$) must be
treated as constructors.
Note that, in a program without overlapping rules, the bodies of two overlapping
facts are forced to be comparable by means of $\sqsubseteq$. 


\begin{definition}[Function {\em clean}]
\label{def_clean}

Given a fact $F$ belonging to an interpretation $I$, let us define the
set $S^F_I=\{F_i \equiv f~ t_{i1} \ldots  t_{in} ~|~ c_i = b_i \in I
~\text{such~that~}F \text{~overlaps~} with~F_i 
\text{~and~}$ $F_i$ \\ $\text{~is~more~specific~than~} F$ 
$(i:1..m)\}$.

Considering the set $S^F_I$ for every fact $F \in I$, we can define
$\mathit{clean}$ (whose type is $\Set{({\cal F})} \rightarrow
\Set{({\cal F})}$) as:

\begin{equation}
\label{eq_clean}
\begin{split}
  &\mathit{clean}(I)= I - I^\bot - I^O \cup\\
  &\bigcup_{F' \equiv (f~t_1 \ldots t_n | c = b) \in I^O}
         \hspace{-3em}[f~t_1  \ldots t_n ~|~ c ~ \wedge
  \bigwedge_{\forall F_i \in S^{F'}_I}
       (\mathit{nunif}((t_1,\ldots,t_n),(t_{i1},\ldots,t_{in})) \vee
       \mathit{not}(c_i))=b] \\
 &\hspace{25em} (F_i \equiv f~ t_{i1} \ldots  t_{in} | c_i = b_i ) 
\end{split}
\end{equation}

\noindent
where $-$ stands for set subtraction and:

\begin{itemize}
\item
$I^\bot=\{l=\sbot ~\text{such~that~} (l=\sbot) \in I\}$. {\em clean}
  removes all the facts that are identically $\bot$.

\item
$I^O=\{F' \in I \text{~such~that~} S^{F'}_I \neq \emptyset\}$. All the
  facts $F'$ in $I$ which are overlapped by some more specific fact are
  removed from $I$ and replaced by the {\em amended} fact shown
  above which does not overlap with any fact in $S^{F'}_I$.

\end{itemize}

\end{definition}

\begin{figure}
    \begin{tabular}{p{4.7cm} c l p{3cm}} 
      \multicolumn{4}{c}{$\mathit{nunif} : T \times T \rightarrow$ Bool} 
        \\ \\
        $\mathit{nunif}(x,t)=\mathit{nunif}(t,x)$ &=& False & $t \in T,
        x \in V$
        \\
        $\mathit{nunif}(k,d)$ &=& True & $k,d \in \DC,
        k \neq d$
        \\
        $\mathit{nunif}(k,k)$ &=& False & $k \in \DC$
        \\
        $\mathit{nunif}((p_1,p_2),(p_3,p_4))$ &=& $\mathit{nunif}(p_1,p_3)
        \vee \mathit{nunif}(p_2,p_4)$ & Tuples
        \\
        $\mathit{nunif}(k(\ldots),k'(\ldots))$ &=& True & $k \neq k'$
        \\
        $\mathit{nunif}(k(p_1,\ldots,p_n),k(p'_1,\ldots,p'_n))$ &=& 
         $\mathit{nunif}(p_1,p'_1)    
         \vee \ldots \vee \mathit{nunif}(p_n,p'_n)$ & $k \in \DC$
        \\
    \end{tabular}
\caption{Lack of unification between patterns: function \emph{nunif}}
\label{fig_nunif}
\end{figure}

\noindent
The function {\em nunif} (Fig. \ref{fig_nunif}) denotes lack of
unification between its arguments. 

Under some conditions {\em clean} will not add new facts to the given
interpretation. This will happen if the guards for the facts under
the big $\cup$ in Eq.~\ref{eq_clean} are unsatisfiable. If the program
under analysis meets certain properties, this is sure to happen. Two
definitions are needed to define those properties:

\begin{definition}[Complete Function Definition]
A function definition for function $f$ written in the core language is
said to be complete if and only if for any well typed, full application $f~t_1
\ldots t_n$ of $f$, where the $t_i$ are terms there is a rule $f~p_1
\ldots p_n | g = r$ that can be used to unfold that application (that
is, there exists a substitution $\sigma$ such that
$\sigma(t_1,\ldots,t_n)=(p_1,\ldots,p_n)$ and $\sigma(g)$ satisfiable).
\end{definition}

\begin{definition}[Productive Rule]
A program rule is said to be productive if at least one fact which is
not equal to the unguarded bottom (\sbot) is generated by unfolding that
rule at some interpretation $I_m$  ($m$ finite). 
\end{definition}

\noindent
{\em clean} will not add new facts if all the function definitions
in the program are complete and all the rules in the program are productive.
The following Lemma states this. Note  that the conditions
mentioned are sufficient but not necessary. 

\begin{lemma}[When Can {\em clean} Drop Facts]
\label{lemma_old_clean}
Let $P$ be a program without overlapping rules, whose function
definitions are all complete and whose rules are all productive. Then:

For every fact $H \equiv f~t_1 \ldots t_n ~|~ c = b \in
I_n$ which is a result of unfolding the rule $\Lambda \equiv f~s_1
\ldots s_n ~|~ g = r$, there exist in $I_{n+1}$ some
facts which are also the result of unfolding $\Lambda$ which cover all
the invocations of $f$ covered by $H$.

The proof for this Lemma can be found in the Appendix.  
\end{lemma} 

We will be using the simplified version of {\em clean} whenever the
program under analysis meets the criteria that have been just mentioned.

To finish this section, let us state a result that justifies why it is
legal to use {\em clean} to remove overlapping facts.

\begin{lemma}[Programs without Overlappings]
\label{lemma_clean_overlappings}
The {\em fixpoint interpretation} (namely, 
$I_\omega=U^\infty_P(I_\bot)$ where $U$ is the unfolding operator that
will be presented later) of any program $P$ without overlapping
rules cannot have overlappings. $I_\bot$ is the empty interpretation. 

The proof for this Lemma can be found in the Appendix.  
\end{lemma}

\subsubsection{Lazy Matching of Facts and Rules}
\label{par_match}

\begin{figure}[t]
    \begin{tabular}{p{3cm} p{0.25cm} p{2.5cm} p{6cm}} 
      \multicolumn{4}{c}{$\mathit{match} : T \times E \rightarrow
      (V \Mypfun E) \times E$}  
    \\
    \\
    $\mathit{match}(x,e)$ & = & $(\{x \leftarrow e\},\lstinline.True.)$
    & $x \in V$
    \\
    $\mathit{match}(t,\sbot)$ & = & $(\{\},\lstinline.False.)$
    \\
    $\mathit{match}(t,(f~e_1 \ldots e_n))$ & = &
    $(\sigma'_h,$
    $c'_h \wedge b'_h)$ & $t \in T, f
    \in \FS \cup \PF, $    
    $\mathit{hnf}((f e_1 \ldots e_n))=c'_h \MyIfThenOp e'_h$.
    $\mathit{match}(t,e'_h)=(\sigma'_h, b'_h)$.
    \\
    $\mathit{match}(k,k)$ & = & $(\{\},\lstinline.True.)$ & $(k \in \DC)$
    \\
    $\mathit{match}((k \ldots),(k' \ldots))$ & = &
    $(\{\},\lstinline.False.)$ & $(k,k' \in \DC, k \neq k')$
    \\
    $\mathit{match}((k~t_1 \ldots t_n),$ 
    \\
    ~~~~~~$(k~e_1 \ldots e_n))$ & = &
    $(\sigma_1 \circ \dots \circ$  
    $\sigma_n,$ \\
    & & ~~~$b_1 \wedge \ldots \wedge $
    $b_n)$ 
    & $(t_i \in T, k \in \DC, \mathit{match}(t_i,e_i)=(\sigma_i,b_i))$
    \\
    $\mathit{match}(t,(f~e_1 \ldots e_n))$ & = &
    $(\{\},\lstinline.False.)$ & $(f \in \FS_m \cup \PF_m, m > n)$ 
    \\
    $\mathit{match}(t,x)$ & = & $(\{x \leftarrow t\},\lstinline.True.)$ & $(t \in
    T, x \in V)$ 
    \\
    \\
    \end{tabular}
\caption{Function \emph{match}.}
\label{fig_match_function}
\end{figure}

The unfolding process {\em rewrites} user-defined
function applications but predefined functions (including partial
application) will be left unaltered by the unfolding steps since there
are no rules for them. This means that when a match is sought to
perform an unfolding step, the arguments to the user-defined
functions may include predefined functions that must be evaluated before
it is known whether they match some pattern. Such applications may
also generate infinite values. Thus, we need a function
{\em match}\footnote{Note that \lstinline.match. is similar to operator
  \lstinline.=:<=. proposed in \cite{Antoy05declarativeprogramming}.}  
 that lazily matches a pattern to an expression.

Recall Fig.~\ref{fig_higher}. The unfolding operator generates facts
containing \lstinline.match. whenever it finds a subexpression headed by a
symbol in $\PF$ that needs to be matched against some rule
pattern. These special facts can be thought as imposing
assumptions on what the pattern must be like before proceeding. 

\begin{figure}[t]
  \begin{tabular}{p{0.5cm} p{1.5cm} p{0.5cm} p{2cm} p{8cm}} 
    \multicolumn{5}{c}{$\mathit{umatch} : T \times E \rightarrow (V \Mypfun E) \times E$}
    \\
    \multicolumn{2}{l}{$\mathit{umatch}(t,e)$} & = & $(\sigma, \lstinline.True.)$ &
   \multicolumn{1}{l}{if there exists some unifying $\sigma$ such that $\sigma(t) \equiv e$}.
    \\
    \multicolumn{2}{l}{$\mathit{umatch}(t,e)$} & = & \multicolumn{2}{l}{$(\sigma, \lstinline.snd(match.(t|_o,e|_o)\lstinline.). \wedge c)$} \\
    & \multicolumn{4}{l}{if $e$ and $t$ do not unify because there is at
       least a position $o$ 
    such that $e|_o$ is headed} 
    \\
    & \multicolumn{4}{l}{by a symbol of $\PF$ 
    (including \lstinline.@.) and $t|_o$ is not a variable.  
    $\mathit{umatch}(t,e[(t|_o)]|_o) =  (\sigma, c).$}
    \\
    \multicolumn{2}{l}{$\mathit{umatch}(t,e)$} & = & $(\epsilon,
    \lstinline.False.)$ &
    if $e$ and $t$ do not unify but this is not due to a predefined
    function symbol in $e$. 
  \end{tabular} 
\caption{umatch: Generation of matching conditions.}
\label{fig_umatch}
\end{figure}

Those assumptions are included inside the fact's guard.   Two functions
are needed in connection to those assumptions: {\em umatch}
(Fig.~\ref{fig_umatch}) 
\footnote{Observe that a function like {\em umatch} is not needed in pure Prolog
since every atom is guaranteed to have a rule and lack of instantiation
will cause a runtime error.}
 generates them as a conjunction of calls to {\em match}
 (Fig.~\ref{fig_match_function}) which performs the matches at runtime.  

{\em umatch} and {\em match} must be distinguished: {\em
  umatch} {\em fits} facts' heads into expressions for unfolding while 
{\em match} is not an unfolding function; it is a 
function used to check (at runtime) whether certain
conditions are met in evaluated expressions. {\em umatch} does not
call {\em match}: {\em umatch} generates code that uses {\em match}. 

The function {\em hnf}, used in the definition for {\em match}, 
receives an expression and returns
that expression evaluated to Head Normal Form. $\mathit{hnf}$ has type
$E \rightarrow E$. 

In the result  of {\em umatch}, $\sigma$ is a list of assignments assigning
values to variables inside the arguments passed to {\em umatch} and
the right part of the result is a condition of the form 
 $\bigwedge_i \mathit{snd}(\mathit{match(p_i,e_i))}$ where the $p_i$
are patterns and the $e_i$ are expressions without symbols of $\FS$ (they have
been removed by unfolding).
 
The function {\em match} returns whether that matching was possible
and a list of assignments from variables to expressions. The rules of
{\em match} are tested from the first to the last, applying the first
suitable one only. 

Both lists of assignments (the ones returned by {\em umatch} or {\em match})
are not exactly substitutions because variables can be 
assigned to full expressions (not just terms) but they behave as
such. 

Two remarks must be made about {\em match}:
(i) The first element of the pair returned by {\em match} is never used
inside the definitions given in this paper because it is only used in
order to bind variables at runtime (not at 
unfolding time). Those bindings will occur when a guard containing
calls to {\em match} is evaluated. (ii) Therefore, {\em match} is not
purely functional (i.e., it is not a side effect-free). 

\begin{example}{(How {\em umatch works}.)}
\label{umatch_example}
Code that generates a situation like the one
described is the one in Fig.~\ref{fig_match} left. Part of
its unfolding appears in Fig.~\ref{fig_match} right \footnote{The
variables in the unfolder's output have been renamed to ease understanding.}.

When the rule for \lstinline.app_first. is unfolded, it is found that
\lstinline.(f n). cannot be unfolded any more but it still does not
match \lstinline.(x:xs). (the pattern in \lstinline.first.'s
rule). Therefore, the second rule for {\em umatch} imposes the
assumption in the resulting fact that \lstinline.(f n). must match
\lstinline.(x:xs). if the rule for \lstinline.app_first. is to be
applied. 
Note that \lstinline.f@[n]. (\lstinline.f. applied to variable \lstinline.n.)
  generates an infinite term in this case. This is why \lstinline.match.
  cannot be replaced by strict equality.
Example \ref{ones_example} in the Appendix (Sect.~
\ref{sec_additional_examples}) shows how unfolding behaves when  
infinite structures are generated.

\begin{figure}[t]
\begin{center}
\begin{tabular}{ p{5cm} | p{7cm} }
{\bf a) Code that Needs Matching}
&
{\bf b) Unfolding of the Source Code}
\\
\begin{lstlisting}
from_n::Int->[Int]
from_n n = n:(from_n(n+1))

first::[a]->a
first (x:xs) = x

app_first :: (a->[b])->a-> b
app_first f n = first(f n)

main::Int->Int
main n=app_first from_n n\end{lstlisting} 
&
\begin{lstlisting}
* first(Cons(x,xs)) = x
* from_n(n) = 
  Cons(n,Cons(n+1,Bot))
* app_first(f,n) | 
  snd(match(Cons(x,xs),f@[n]))=x 
\end{lstlisting}
{\bf Note:} Any code preceded by \texttt{*} in every line has been 
generated by our Prolog-based unfolder. The unfolder uses Prolog terms
to represent functional applications. That is why the
unfolder uses tuples to represent curried applications.
\end{tabular}
\end{center}
\vspace{-3ex}
\caption{Lazy matching of terms and rules.}
\label{fig_match}
\end{figure}

\end{example}

\subsubsection{Unfolding Operator}
\label{par_immediate}
Operator $U(I)$ (short form for $U_P(I)$) where $I$ is a $\pi$-interpretation is
defined as shown in Fig.~\ref{fig_unfolding_operator}. 

\begin{figure}[h]
$I_0=I^\bot_0=\emptyset$
\\
$I_{m+1}=U(I_m)=\mathit{clean}(I_{m+1}^\top)$
\\
$I^\top_{m+1}=\bigcup_{\Lambda \in 
     \mathit{Rules}}(\mathit{unfold}(\Lambda,I_m \cup I^\bot_m)) \cup
   I_m$
\\
$I^\bot_{m+1}=\{l=\sbot ~\text{such~that~} (l=\sbot) \in I^\top_{m+1}\}$
\caption{Unfolding operator}
\label{fig_unfolding_operator}
\end{figure}

Given a program P, its meaning is given by the least fixed point of
$U_P$ or by $I_\omega (=U^\infty_P(I_\bot))$ if the program has
infinite semantics. 

The auxiliary function {\em unfold}, that unfolds a rule using the
facts in an interpretation, is defined in Fig.~\ref{fig_unfold}. 
The behaviour of {\em unfold} can be described as follows: {\em
unfold} receives a (partially) unfolded rule (a {\em pseudofact})
which is unfolded by means of recursive calls. When the input to {\em
unfold} has no invocations of user defined functions, it is just
returned as it is (case 1). Otherwise, the pseudofact is unfolded by
considering all the facts and positions $o$ which hold an invocation
of a user-defined function (Case 2a). Those positions occupied by
user-defined function calls which cannot be unfolded are replaced by
$\sbot$ (case 2b). {\em unfold} returns all the possible facts
obtained by executing this procedure.

\begin{figure}[t]
  \begin{center}
  $\mathit{unfold} : \mathit{Rule} \times \Set({\cal F}) \rightarrow
    \Set({\cal F})$
  \end{center}

  $\mathit{unfold}(l ~|~ g = r,I_m)=$ \\  
  \[
     \begin{cases}
     \{l~|~\mathit{eval}(I_m,g) = \mathit{eval}(I_m,r) \} ~
         \text{if $g$ and $r$ have no total apps. of user funcs.} 
  \\
  \\
  \{(h''~|~c''=b'') \text{~such~that~} (h''~|~c''=b'') \in 
     &
     \text{A position is unfoldable:}
      \\
      ~~\mathit{unfold}(\sigma(l)~|~\mathit{eval}(I_m,\sigma(g\langle
      b_j\rangle\|_o \wedge c_j \wedge c'_m)) =
      \mathit{eval}(I_m,\sigma(r\langle b_j \rangle\|_o)),I_m)
     &
     \text{Case 2a):} 
      \\
      ~~~~\forall o \in \mathit{Pos}(r) \cup \mathit{Pos}(g),
      r|_o~(\mathit{resp.~g|_o})~=f~e_1 \ldots e_n, f \in \FS_n
      & \text{Some facts {\em fit} position $o$} 
      \\
      ~~~~\forall (f~t_1 \ldots t_n~|~c_j=b_j) \in I_m~\text{such~that}
      \\
      ~~~~~~\mathit{umatch}((t_1,\ldots,t_n),(e'_1,\ldots,e'_n))=(\sigma,c'_m)
      ~\text{and~$c'_m$~satisfiable}
      \\
      ~~\cup
      \\
      ~~\mathit{unfold}(l~|~\mathit{eval}(I_m,g\langle \sbot \rangle\|_o) =
      \mathit{eval}(I_m,r\langle \sbot \rangle\|_o),I_m)
      &
      \text{Case 2b):}
      \\
      ~~~~\forall o \in \mathit{Pos}(r) \cup \mathit{Pos}(g),
      r|_o~(\mathit{resp.~g|_o})~=f~e_1 \ldots e_n, f \in \FS_n
      &
      \text{No facts {\em fit} position $o$}
      \\
      ~~~~\nexists (f~t_1 \ldots t_n~|~c_j=b_j) \in I_m~\text{such~that}
      \\
      ~~~~~~\mathit{umatch}((t_1,\ldots,t_n),(e'_1,\ldots,e'_n))=(\sigma,c'_m)
      ~\text{and~$c'_m$~satisfiable}
 \}
     \end{cases}
  \]
  
where:

\begin{itemize}
\item
$e_i'=\mathit{eval}(I_m,e_i) ~~~ \forall i:1..n$

\item
$g\langle t \rangle \|_o=g[t]|_o$ if $o \in \mathit{Pos(g)}$ and
  $g\langle t \rangle \|_o=g$ otherwise. 

\item
$r\langle t \rangle
  \|_o=r[t]|_o$ if $o \in \mathit{Pos(r)}$ and 
  $r\langle t \rangle \|_o=r$ otherwise. 
\end{itemize}

\caption{Unfolding of a program rule using a given interpretation}
\label{fig_unfold}
\end{figure}

When performing the unfolding of a program, {\em unfold}
behaves much like the rewriting process in a TRS (i.e., it tries all
the possible pairs $\langle$position $o$, fact$\rangle$).

\noindent
To summarize, $\sbot$ and \lstinline.match. are the two enhancements
required to write valid code for unfolding functional programs.
If eager evaluation is used, these enhancements would not be
necessary but naïve unfolding would still fail to work.

\section{Operational Semantics}

The operational semantics that describes how
ground expressions written in the kernel language are evaluated is
shown in Fig.~\ref{fig_operational_exp}.  
The semantics defines a small step relationship denoted by
$\leadsto$. The notation $e \leadsto e'$ means that the expression $e$
can be rewritten to $e'$.
The reduction relation $(p~e_1 \ldots e_n) \leadsto^p t ~~(p \in
\PF)$ states that $p~e_1 \ldots e_n$ can be rewritten to $t$ by using the definition
of the predefined function $p$.



The unfolding and operational semantics are equivalent in the
following sense for any ground expression $\mathit{goal}$:
\fbox{$\mathit{goal} \leadsto^* e' \leftrightarrow
e' \in \mathit{ueval}(I_\infty,\mathit{goal}) $}
where $\leadsto^*$ is the transitive and reflexive closure of
$\leadsto$ and $e'$ is in normal form according to $\leadsto$,
$\mathit{ueval}$ is a function that evaluates expressions by means of
unfolding and $I_\infty$ is the {\em limit} of the interpretations
found by repeatedly unfolding the program. This equivalence is proved
in the Appendix, Sect.~\ref{sec_equivalence}.

\begin{figure}[t]
\begin{center}
\begin{tabular}{m{5.4cm} p{1cm} m{6cm} p{0.6cm}}
\[
 \frac{
   \begin{split}
   e=f~e_1 \ldots e_n, f \in \FS_n, \\[-3pt]
   (f~t_1 \ldots t_n|g=r) \in P \\[-3pt]
   \sigma=\mathit{mgu}((t_1,\ldots,t_n),(e_1,\ldots,e_n)), \\[-3pt]
   \sigma(g) \leadsto^*\mathit{True}
   \end{split}
 }
 {
   e \leadsto \sigma(r)
 }
\]
&
\multirow{1}{*}{\textsc{(rule)}}
& 
\[
 \frac{
   \begin{split}
   e=f~e_1 \ldots e_n, f \in \FS_n, \\[-3pt]
   \nexists (\Lambda \in P \text{~such~that~} \Lambda \equiv (f~t_1 \ldots t_n|g=r) \\[-3pt]
   \sigma=\mathit{mgu}((t_1,\ldots,t_n),(e_1,\ldots,e_n)), \\[-3pt]
   \sigma(g) \leadsto^*\mathit{True})
   \end{split}
 }
 {
   e \leadsto \sbot
 }
\]
&
\multirow{1}{*}{\textsc{(rulebot)}}
\\[-20pt]
\[
 \frac{
   e=(p~e_1 \ldots e_n), e \leadsto^p t, t \in T, p \in \PF
 }
 {
   e \leadsto t
 }
\]
&
\multirow{1}{*}{\textsc{(predef)}}
& 
\begin{center}
  $\mathit{True} \wedge \mathit{True} \leadsto \mathit{True}$
\end{center}
&
\multirow{1}{*}{\textsc{(andtrue)}}
\\[-20pt]

  \begin{tabular}{m{2.8cm} p{1.5cm} m{3cm} p{2cm} m{2.5cm} p{1cm}}
  \[
    \frac{
      e_i \leadsto^* \mathit{False}, i:1,2
    }{
      e_1 \wedge e_2 \leadsto \mathit{False}
    }
  \]
  &
  \multirow{1}{*}{\textsc{(andfalse)}}
  & 
  ~~~~$(\mathit{False} \MyIfThenOp e) \leadsto \sbot$
  &
  \multirow{1}{*}{\textsc{(ifthenfalse)}}
  &
  ~~~~$(\mathit{True} \MyIfThenOp e) \leadsto e$
  &
  \multirow{1}{*}{\textsc{(ifthentrue)}}
  \end{tabular}
\end{tabular}
\end{center}
\caption{Operational Semantics}
\label{fig_operational_exp}
\end{figure}

Note that this semantics is fully indeterministic; it is not meant to
be used in any kind of implementation and its only purpose is to serve
as a pillar for the demonstration of equivalence between the unfolding
and an operational semantics. Therefore, the semantics is not lazy or
greedy in itself. It is the choice of reduction positions where the
semantics' rules are apllied what will make a certain evaluation lazy or not.

\section{Some Applications of the Unfolding Semantics}
\label{sec_applications}




\paragraph{Declarative Debugging}\footnote{The
    listings of unfolded code provided in this paper have been
    generated by our {\em unfolder}. Source at
    \url{http://www.github.com/josem-rey/unfolder} and test
    environment at
    \url{https://babel.ls.fi.upm.es/~jmrey/online_unfolder/unfolding.html}}
With declarative debugging, the debugger {\em consults} the internal
structure of source code to find out what expressions depend on other
expressions and turns this information into an Execution Dependence
Tree (EDT). The debugger uses this information as well as answers from
the user to blame an error on some rule.
We have experimentally extended the unfolder to collect intermediate results
as well as the sequence of rules that leads to every fact. This
additional information allows our unfolder to build the EDT for any
program run. Consider for example this buggy addition:

\lstset{style=bluecode}
\begin{lstlisting}
A1: addb Zero n = n
A2: addb Suc(Zero) n = Suc(n)
A3: addb Suc(Suc(m)) n = Suc(addb m n)
M24: main24 = addb Suc(Suc(Suc(Zero))) Suc(Zero)
\end{lstlisting}   
\lstset{style=code}

\noindent
We can let the program unfold until \lstinline.main24. is fully
evaluated. This happens in $I_3$, which contains the
following fact for the main function (after much formatting):

\lstset{style=bluecode}
\begin{lstlisting}
root:main24 = Suc(Suc(Suc(Zero))) <M24>
  n1:addb(Suc(Suc(Suc(Zero))),Suc(Zero))=Suc(Suc(Suc(Zero)))<A3>
    n2:addb(Suc(Zero),Suc(Zero)) = Suc(Suc(Zero)) <A2>
\end{lstlisting}
\lstset{style=code}

\noindent
Now, following the method described in \cite{Pope98buddha-}, we can 
think of the sequence above as a 3-level EDT in which the root and
node \lstinline.n1. contain wrong values while the node
\lstinline.n2. is correct, putting the blame on rule A3.

The main reason that supports the use of unfolding for performing
declarative debugging is that it provides a platform-independent
environment to test complex programs. This platform independence can
help check the limitations of some implementations (such of unreturned
answers due to endless loops).

\paragraph{Test Coverage for a Program}
It is said that a test case for a program {\em covers} those rules
that are actually used to evaluate the test case.
We would like to reach full code coverage with the
smallest test set possible. The unfolder can be a valuable tool for finding
 such a test set if it is enhanced to record the
list of rules applied to reach every fact.  

What must be done with the enhanced unfolder is to calculate
interpretations until all the rules appear at least once in the rule
list associated to the facts that do not contain any $\sbot$ and
then apply a minimal set coverage algorithm to find the set of facts
that will be used as the {\em minimal} test set.
For example:

\lstset{style=bluecode}
\begin{lstlisting}
R1: rev [] = []    // List inversion
R2: rev (x:xs) = append (rev xs) [x]
A1: append [] x = x
A2: append (x:xs) ys = x:(append xs ys)
\end{lstlisting}
\lstset{style=code}

\noindent
The first interpretation contains:

\lstset{style=bluecode}
\begin{lstlisting}
* rev(Nil) = Nil  <R1>
* append(Nil,b) = b  <A1>
* append(Cons(b,c),d) = Cons(b,Bot)  <A2>
\end{lstlisting}
\lstset{style=code}

\noindent
So, appending the empty list to any other achieves 50\%
coverage of \lstinline.append..  Reversing the empty list
uses 1 rule for \lstinline.rev.: the coverage rate is 50\% too. 
$I_3$ has: 

\lstset{style=bluecode}
\begin{lstlisting}
* append(Cons(b,Nil),c) = Cons(b,c) <A2,A1>
...
* rev(Cons(b,Cons(c,Nil))) = Cons(c,Cons(b,Nil)) 
  <R2,R2,R1,A1,A2,A1>
\end{lstlisting}
\lstset{style=code}

\noindent
This shows that the {\em minimal} test set to test
\lstinline.append. must consist of appending a one element list to any
other list. Meanwhile, reversing a list with 2
elements achieves a 100\% coverage of the code: all the rules are
used.  

To close this section, we would like to mention that Abstract
Interpretation can be used along with unfolding  
to find properties of the programs under study such as
algebraic or demand properties. See examples \ref{ex_abs1},
\ref{ex_parities_revisited}, \ref{ex_demand_analysis} in the Appendix.

\section{Conclusion and Future Work}

We have shown that unfolding can be used as the basis for the
definition of a semantics for lazy, higher-order functional programs
written in a kernel language of conditional equations. This is done by
adapting ideas from the \emph{s}-semantics approach for logic
programs, but dealing with the aforementioned features was not trivial,
and required the introduction of two ad-hoc primitives to the
kernel language: first, a syntactic representation of the undefined
and second, a matching operator that deals with partial information.

Effort has also been devoted to simplifying the 
code produced by the unfolder, by erasing redundant facts and constraining the
shape of acceptable facts. We have provided a set of requirements for
programs that ensure the safety of these simplification procedures. 
We have also proven the equivalence of the proposed unfolding
semantics with an operational semantics for the kernel language.

We have implemented an unfolder for our kernel
language. Experimenting with it supports our initial claims about a
more ``implementable'' semantics.  

Regarding future work, we want to delve into the applications that
have been just hinted here, particularly declarative debugging and
abstract interpretation. 

Finally, we are working on a better characterization of the
\emph{necessary} conditions that functional programs must meet in
order for different optimized versions of the \emph{clean} method to
work safely.

%
%


\bibliographystyle{splncs03}

\bibliography{unfolding}

\clearpage

\appendix

\section*{APPENDIX}

This appendix is \emph{not} part of the submission itself and is
provided just as supplementary material for reviewers.  It pursues
the following goals:

\begin{enumerate}
\item
To provide a pictorical representation of the functions involved in
the unfolding process, which hopefully helps in grasping how the whole
process works (Sect.~A).

\item
To describe in what sense the unfolding and the operational semantics
are equivalent and to prove such equivalence (Sect.~B).

\item
To present a larger example that intends to clarify how the functions
that have been used actually work as well as additional examples
(Sect.~C). 

\item
To establish some results that support the validity of
the code generated by the unfolder (Sect.~D).
\end{enumerate}

\section{Pictorial Representation of the Unfolding Process}
\label{sec_unfolding_pictorial}

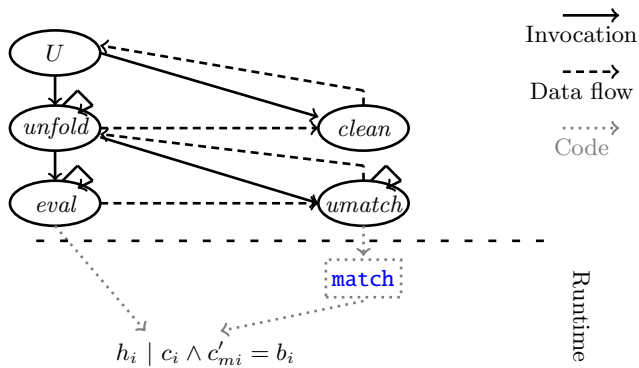
\begin{figure}[h]

\begin{tikzpicture}[line width=1pt]

\draw (0,1.5) ellipse (.6 and .3);
\draw (0,1.5) node {$U$};
\draw [->] (0.6,1.5) -- (3.5,0.65);
\draw (4.1,0.5) ellipse (.6 and .3);
\draw (4.1,0.5) node {{\em clean}};
\draw [->] (0,1.2) -- (0,0.8);
\draw (0,0.5) ellipse (.6 and .3);
\draw (0,0.5) node {{\em unfold}};
\draw [->] (0,0.2) -- (0,-0.2);
\draw (0,-0.5) ellipse (.6 and .3);
\draw (0,-0.5) node {{\em eval}};
\draw [->] (0.6,0.38) -- (3.5,-0.5);
\draw (4.1,-0.5) ellipse (.6 and .3);
\draw (4.1,-0.5) node {{\em umatch}};
\draw [gray,dotted,->] (0,-0.8) -- (1.2,-2.2);
\draw [loosely dashed] (-0.25,-1) -- (6.5,-1);
\draw [gray,dotted,->] (4.1,-0.8) -- (4.1,-1.2);
\draw [gray,dotted] (3.6,-1.25) rectangle (4.6,-1.75);
\draw [gray] (4.1,-1.5) node {{\lstinline.match.}};
\draw [gray,dotted,->] (4.1,-1.8) -- (2.2,-2.2);
\draw (2,-2.5) node {\small $h_i~|~c_i \wedge c'_{mi}=b_i$};

\draw (0.1,0.8) -- (0.3,1.0);
\draw (0.3,1.0) -- (0.5,0.8);
\draw [->] (0.5,0.8) -- (0.3,0.7);

\draw (0.1,-0.2) -- (0.3,0.0);
\draw (0.3,0.0) -- (0.5,-0.2);
\draw [->] (0.5,-0.2) -- (0.3,-0.3);

\draw (4.2,-0.2) -- (4.4,0.0);
\draw (4.4,0.0) -- (4.6,-0.2);
\draw [->] (4.6,-0.2) -- (4.4,-0.3);

\draw [densely dashed,->] (0.6,0.5) -- (3.5,0.5);

\draw [densely dashed,->] (0.6,-0.5) -- (3.5,-0.5);

\draw [densely dashed] (4.1,-0.2) -- (4.1,0.0);
\draw [densely dashed,->] (4.1,0.0) -- (0.6,0.42);

\draw [densely dashed] (4.1,0.8) -- (4.1,1.0);
\draw [densely dashed,->] (4.1,1.0) -- (0.6,1.6);

\draw (7,-2) node [rotate=270] {Runtime};

\draw [->] (6.75,2) -- (7.5,2);
\draw (7,1.75) node {Invocation};

\draw [->,densely dashed] (6.75,1.25) -- (7.5,1.25);
\draw (7,1) node {Data flow};

\draw [gray,->,dotted] (6.75,0.5) -- (7.5,0.5);
\draw [gray] (7,0.25) node {Code};
\end{tikzpicture}

\caption{Relation among the functions taking part in unfolding}
\label{fig_overview_unfolding}
\end{figure}

\noindent
Throughout Sect.~\ref{sec_unfolding_sem} a number of auxiliary functions were
presented. These functions are depicted in
Fig.~\ref{fig_overview_unfolding}. The figure can be explained as follows:

The starting point is $U$. $U$ does nothing but to call {\em unfold}
and remove the {\em redundant} facts by calling {\em clean}. It is
then up to the user to call $U$ again to perform another step in the
unfolding process.

The second level of the figure shows {\em unfold}, which takes a
program rule and unfolds it as much as possible. {\em unfold} calls
itself with the output of its previous execution until no more
positions are left to unfold (arrow pointing downwards). If {\em
unfold} receives an input where at least one position is unfoldable,
it calls {\em eval} on the arguments of the unfoldable expression
and then calls {\em umatch} to perform the actual {\em fitting}
between the unfoldable position and the head of some fact.

The last level of the figure (below the dashed line) represents the
execution of the unfolded code. This part is not related with the
definition of the unfolding operator, but with the execution of the
unfolded code. The code is made of the output of {\em unfold} whose
guards are (possibly) extended with $c'_{mi}$, the output from {\em
  umatch}, which contains the invocations to {\em match}. Observe that
the output from {\em umatch} goes to the generated code only, not to
the unfolding process.     

To the best of our knowledge, this unfolding process is a first effort
to formulate an unfolding operator beyond naïve unfolding.

\section{Equivalence between the Unfolding Semantics and the
  Operational Semantics}

\subsection{Unfolding of an Expression}
\label{sec_ueval}

Let us define a function $\mathit{ueval}$ that finds what is the
normal form for a given expression by means of unfolding. In short,
what {\em  ueval} does is to evaluate a  
given (guarded) expression by unfolding it according to a given
interpretation. 

The function {\em ueval} has type $\mathit{ueval} :
\Set({\cal F}) \times E \rightarrow \Set(E)$ and is defined as shown
in Fig.~\ref{fig_ueval}. 

\begin{figure*}[t]
  \begin{tabular}{p{8cm} p{0.4cm} p{5cm}}
     \multicolumn{2}{c}{$\mathit{ueval} : \Set({\cal F}) \times
     E \rightarrow \Set(E)$} \\
     \\
     $\mathit{ueval}(I,e)$ & = & $\{e\}$ ~~if no rule from
       $\mathit{evalAux}$ applies to any position of $e$\\
     $\mathit{ueval}(I,e)$ & = & $\bigcup_o
       \mathit{ueval}(I,e[\mathit{uevalAux}(I,e|_o)]|_o)$ ~~$\forall
       o$~such that a rule of $\mathit{uevalAux}$ is applicable to $e|_o$. \\
     \multicolumn{3}{c}{- - - - - - - - - - - - -} \\
  \end{tabular}

  \begin{tabular}{p{0.5cm} p{8cm} p{0.4cm} p{7.5cm}}
     \\
     \multicolumn{2}{c}{$\mathit{uevalAux} : \Set({\cal F}) \times E \rightarrow E$} \\
    \\
    \multicolumn{2}{l}{$\mathit{uevalAux}(I,p~e_1 \ldots e_n)$} & = & t \\
      & \multicolumn{3}{l}{\text{if $p~e_1 \ldots e_n \leadsto^p t ~~(p \in \PF_n)$}}\\
  \\
    \multicolumn{2}{l}{$\mathit{uevalAux}(I,(c_1 \wedge \ldots \wedge c_{i-1} \wedge \mathit{snd}(\mathit{match}(p,e)) \wedge
    c_{i+1} \wedge \ldots \wedge c_n \MyIfThenOp e'))$} & = & $\sigma(c_1 \wedge c_{i-1} 
    \wedge b \wedge c_{i+1} \wedge \ldots \wedge
    c_n \MyIfThenOp e')$ \\
      & \text{if $\mathit{match}(p,e)=(\sigma,b)$}\\
  \\
    \multicolumn{2}{l}{$\mathit{uevalAux}(I,\mathit{True} \wedge \mathit{True})$} & = & $\mathit{True}$ \\
  \\
    \multicolumn{2}{l}{$\mathit{uevalAux}(I,b_1 \wedge b_2)$} & = & $\mathit{False}$ \\
      & $b_1, b_2 \in \mathit{Bool}, b_1=\mathit{False}~\text{or}~b_2=\mathit{False}$
  \\
  \\
    \multicolumn{2}{l}{$\mathit{uevalAux}(I,\mathit{True} \MyIfThenOp e)$} & = & $e$ 
  \\
  \\
    \multicolumn{2}{l}{$\mathit{uevalAux}(I,\mathit{False} \MyIfThenOp e)$} & = & $\sbot$ 
  \\
  \\
    \multicolumn{2}{l}{$\mathit{uevalAux}(I,f~e_1 \ldots e_n)$} & = &
  $\sigma'(c \MyIfThenOp b)$ \\
      & \multicolumn{3}{l}{\text{if} $f \in \FS_n, \exists f~t_1 \ldots t_n|c=b \in I.$ 
      $\sigma'=\mathit{mgu}((t_1,\ldots,t_n),(e_1,\ldots,e_n))$
  with $\sigma'(c)=\mathit{True}$}\\
  \\
  \\
    \multicolumn{2}{l}{$\mathit{uevalAux}(I,f~e_1 \ldots e_n)$} & = &
  $\sbot$ \\
      & \multicolumn{3}{l}{\text{if} $f \in \FS_n, \nexists f~t_1 \ldots t_n|c=b \in I.$ 
      $\sigma'=\mathit{mgu}((t_1,\ldots,t_n),(e_1,\ldots,e_n))$
  with $\sigma'(c)=\mathit{True}$}\\
\end{tabular}

\caption{The $\mathit{ueval}$ function: Evaluating expressions by means
  of unfolding}
\label{fig_ueval}
\end{figure*}

\noindent
Note that any expression $e$ is equivalent to ($\mathit{True}
\MyIfThenOp e$).

\subsection{Trace of a Fact or an Expression}
\label{sec_trace}

Given a fact $F$, belonging to any interpretation $I$, its trace is
the list of pairs $(\Lambda_i,o)$ where 
$\Lambda_i$ is a rule in $\mathit{rule}(P) \cup
\{\Lambda_{\sbot,f}=f~x_1 \ldots x_n = \sbot\} ~~ \forall f \in \FS_n$
and $o$ is a position within the 
expression to which the next rule in the trace is to be applied. This
position indicates what subexpression within the current expression is
to be replaced by the body of the rule applied. 

Let us define the function that returns all the traces
associated to all the facts derivable from a single rule (++ denotes
list of lists concatenation that returns the list of lists resulting
from appending every list in the first argument to every list in the
second argument):

\[
  \mathit{tr'} : \Set({\cal F}) \times R \rightarrow [\tau]
\]

\noindent
where $R$ is the type of program rules and $\tau$ is the type of traces. 

\begin{itemize}
\item
$tr'(I,f~t_1 \ldots t_n|c=b)=[~]$ if $f~t_1 \ldots t_n|c=b$ is a valid
  input for the case 1 of $\mathit{unfold}$.

\item
$tr'(I,f~t_1 \ldots t_n|c=b)= o . \mathit{tr}(I,F) ++
  \mathit{tr'}(I,F')$ if the case 2a) of $\mathit{unfold}$ can be
  applied to $f~t_1 \ldots t_n|c=b$ using fact $F \in I$ at
  position $o$. $F'$ is the result of unfolding $f~t_1 \ldots t_n|c=b$
  as it is done in the aforementioned case of $\mathit{unfold}$.

\item
$tr'(I,f~t_1 \ldots t_n|c=b)=[[(\Lambda_{\sbot,f'},o)]]++
  \mathit{tr'}(I,F')$ if the case 2b) of $\mathit{unfold}$ can be
  applied to $f~t_1 \ldots t_n|c=b$ at position $o$. $F'$ is
  the result of unfolding $f~t_1 \ldots t_n|c=b$ as it is done in 
  case 2b) of $\mathit{unfold}$. 

\end{itemize}

\noindent
The composition of a position $o$ and a trace (denoted by $o
. \mathit{tr}(\ldots)$ above) is defined (for every trace in a given
list) as:  

\begin{itemize}
\item
$o . [~] = [~]$

\item
$o . [(\Lambda,o') | \mathit{xs}] = [(\Lambda,o.o') | o . \mathit{xs}]$
\end{itemize}

\noindent
The list of traces for a fact $F$ with respect to an interpretation $I$
($\mathit{tr}(I,F)$) relies on $\mathit{tr'}$:

\[
  \begin{split}
      \mathit{tr} : \Set({\cal F}) \times {\cal F} \rightarrow [\tau] \\
      \mathit{tr}(I,F)=[[(\Lambda_F,\{\})]] ++ \{\tau \text{~such that~}
      \tau \in \mathit{tr'}(I,\Lambda_F) \wedge \\
      \mathit{unfold}(\Lambda_F,I) \text{~generates~} F
      \text{~according to the steps given by the trace~} \tau \}
  \end{split}
\]

\noindent
where $\Lambda_F$ is the only program rule that can generate $F$.

Note that $\mathit{tr}$ and $\mathit{tr'}$ are mutually recursive.

The list of traces of an expression $e$ according to interpretation $I$
(denoted $\mathit{Tr}(I,e)$) is defined as the tail of all the lists
in $\mathit{tr}(I,\mathit{goal'}=e)$ where $\mathit{goal'}$ is a new
function name that does not appear in the program $P$ and the
$\mathit{tail}$ of a list is the same list after removing its first
element.

\subsection{Equivalence between the Unfolding and Operational Semantics}
\label{sec_equivalence}

This section will show that the unfolding semantics and the
operational semantics are equivalent in the following
sense for any ground expression $\mathit{goal}$:

\begin{equation}
\begin{split} 
\mathit{goal} \leadsto^* e' \leftrightarrow
e' \in \mathit{ueval}(I_\infty,\mathit{goal}) ~~ \\
\label{eq_unfolding_operational}
\end{split}
\end{equation}

\noindent
where $\leadsto^*$ is the transitive and reflexive closure of
$\leadsto$ and $e'$ is in normal form according to $\leadsto$. 

Given a program $P$, $I_\infty$ is the limit of the following sequence:

\begin{itemize}
\item
$I_0=\emptyset$

\item
$I_{n+1} ~~(n \geq 0) = \bigcup_{\Lambda \in \mathit{rule}(P)}\mathit{unfold}(I_n,\Lambda)$
\end{itemize}

\subsection{Proof of Equivalence}

We are now proving that Eq.~\ref{eq_unfolding_operational} holds.

$\rightarrow)$

\noindent
This part of the double implication will be proven by induction on the
number of $\leadsto$-steps that an expression requires to reach normal
form. 

{\bf Base case (n=0):} If $\mathit{goal} \leadsto^0 e'$, then
$\mathit{goal}=e'$, which means that $\mathit{goal}$ is in normal form
already. Therefore, $\mathit{goal}$ has no full applications of
symbols in $\PF \cup \FS$. In that case,
$\mathit{ueval}(I,\mathit{goal})=\{\mathit{goal}\}~~ \forall I \in
\Set({\cal F})$.  
 
{\bf Induction step:}

Let us take as induction hypothesis that any expression
$\mathit{goal}$ such that 
$\mathit{goal} \leadsto^n e'$ (where $e'$ is in normal form) then
$e' \in \mathit{ueval}(I_\infty,\mathit{goal})$.

Let $e^{n+1}$ be an expression that requires $n+1$ $\leadsto$-steps in
order to reach normal form. Then there must exist (at least) one
expression $e^n$ such that:

\[
  e^{n+1} \leadsto e^n \leadsto^n e'
\]

\noindent
where $e'$ is in normal form. Now, if we prove that both $e^{n+1}$ and
$e^n$ unfold to the same values (that is, $\mathit{ueval}(I_\infty,e^{n+1})=\mathit{ueval}(I_\infty,e^n)$), then we can apply the induction
hypothesis to $e^n$ to state that $e^n \leadsto^n e' \rightarrow
e' \in \mathit{ueval}(I_\infty,e^{n+1})=\mathit{ueval}(I_\infty,e^n)$.

Let us check all the rules in the operational semantics for the single
$\leadsto$ step going from $e^{n+1}$ to $e^n$.

\underline{Rule \textsc{rule}}

In this case, $e^{n+1}= f~e_1 \ldots e_a ~~(f \in \FS_a)$ and
$e^n=\sigma(r)$ (assuming that the rule for $f$ within
program $P$ is $\Lambda \equiv f~t_1 \ldots t_a|g = r$ and
$\sigma=\mathit{mgu}((t_1,\ldots,t_n),(e_1,\ldots,e_n))$). 

By {\em reductio ad absurdum} let us assume now that
$\mathit{ueval}(I_\infty,e^{n+1}) \neq \mathit{ueval}(I_\infty,e^n)$. Then,

\[
\begin{split}
\mathit{ueval}(I_\infty,f~e_1 \ldots e_a) \neq 
\mathit{ueval}(I_\infty,\sigma(g \MyIfThenOp r)) 
\end{split}
\]

\noindent
However, note that $\sigma(\Lambda)$ is equal to the rule instance
$f~e_1 \ldots e_a | \sigma(g) = \sigma(r)$, which
states exactly the opposite of the equation above. We have reached a
contradiction, which means that our initial hypothesis (namely,
$\mathit{ueval}(I_\infty,e^{n+1}) \neq \mathit{ueval}(I_\infty,e^n)$) is false.

\underline{Rule \textsc{rulebot}}

In this case, $e^{n+1}=f~e_1 \ldots e_a ~~(f \in \FS_a)$ and
$e^n=\sbot$. If there is no rule in $P$ whose pattern can unify with
$e^{n+1}$ while at the same time having a satisfiable guard, it is
sure that no fact in any interpretation derived from $P$ will be such
that its head unifies with $e^{n+1}$ while at the same time having a satisfiable
guard (which forces $\mathit{uevalAux}$ to use its last case). That
means that $e^{n+1}$ cannot be reduced to anything 
different from $\sbot$. The same happens with $e^n$ (which is already
equal to $\sbot$). Therefore, $\mathit{ueval}(I_\infty,e^{n+1}) =
\mathit{ueval}(I_\infty,e^n)$ as we wanted to prove.

\underline{Rule \textsc{predef}}

In this case, $e^{n+1}=p~e_1 \ldots e_a$ and $e^n=t~~(t \in T)$ where $ p\in \PF$ and $p~e_1 \ldots e_a$ is has value $t$
according to the predefined functions known to the environment being
used.

Also in this case $\mathit{ueval}(I,p~e_1 \ldots e_a)=\{t\}$ and
$\mathit{ueval}(I,t)=\{t\}$ for any interpretation $I$. This
case simply evaluates predefined functions. 




\underline{Rule \textsc{andfalse}}

In this case, $e^{n+1}=e_1 \wedge e_2$ and $e^n=\mathit{False}$ when
either $e_1 \leadsto^* \mathit{False}$ or $e_2 \leadsto^*
\mathit{False}$. Let us assume without loss of generality that $e_1
\leadsto^* \mathit{False}$.

Since $e^{n+1}$ requires $n+1$ $\leadsto$-steps to reach normal form,
then $e_1$ must take at most $n$ steps to reach its normal form. This
means that the induction hypothesis is applicable to $e_1$ and
therefore $\mathit{ueval}(I_\infty,e_1) \supset \{\mathit{False}\}$. This
in turn means that  
 $\mathit{ueval}(I_\infty,e_1 \wedge e_2) \supset \{\mathit{False}\}$ as we wanted
to prove (assuming that the logical 
connector $\wedge$ is defined as lazy conjunction in $\mathit{eval}$).

The remaining rules (\textsc{andtrue, ifthentrue, ifthenfalse}) are
proven in a similar way. 

Let us proceed now to the reverse implication.

$\leftarrow)$

The proof will be driven by structural induction on the shape of the
expression to be evaluated ($\mathit{goal}$).

Let $\mathit{goal}$ be an expression that has no full applications of
any symbol of $\FS$ or $\PF$. Then,
$\mathit{ueval}(I_\infty,\mathit{goal})=\{\mathit{goal}\}$ and
$\leadsto$ cannot apply 
any rewriting, so $\mathit{goal} \leadsto^* \mathit{goal}$, as we
wanted to prove.

Next, let $\mathit{goal}=p~e_1 \ldots e_n ~~ (p \in \PF_n)$ and no
$e_i$ has any full application of any symbol in $\FS \cup \PF$. Then,
$\mathit{ueval}(I_\infty,\mathit{goal})=\{t\}$ if $\mathit{goal}
\leadsto^p t$. $\leadsto$ will apply \textsc{predef,
ifthentrue, ifthenfalse, andtrue} or \textsc{andfalse} to
evaluate the same predefined function and reach the same $t$.

Next, let $\mathit{goal}=f~e_1 \ldots e_n ~~ (f \in \FS_n)$. If
$\mathit{ueval}(I_\infty,\mathit{goal})$ includes $e'$ that is because
$\mathit{goal}$ has a trace (since $e'$ is in normal form) That is,
$[(\Lambda_1,o_1),\ldots,(\Lambda_k,o_k)] \in
\mathit{Tr}(I_\infty,\mathit{goal})$ 
for some $k>0$. We are now going to prove that: 

\[
  \begin{split}
  e' \in \mathit{ueval}(I_\infty,\mathit{goal}) \wedge 
   [(\Lambda_1,o_1),\ldots,(\Lambda_k,o_k)] \in
   \mathit{Tr}(I_\infty,\mathit{goal}) \\
  \rightarrow \mathit{goal} \leadsto^* e' \text{(using the exact
    sequence of rules given below)}
  \end{split}
\]

\noindent
Specifically, it will be proven that every trace element $(\Lambda,o)$
is equivalent to the following $\leadsto$-sequence at position $o$ of
the expression input for the trace element:

\begin{enumerate}
\item
The rules dealing with predefined functions (namely, \textsc{predef,
  ifthentrue, ifthenfalse, andtrue, andfalse}) will be applied to the
expressions being rewritten as many times as possible.

\item
\textsc{rule} or \textsc{rulebot}: The rule \textsc{rule} will be applied if
$\Lambda \neq \Lambda_{\sbot,f}$. \textsc{rulebot} will be applied
otherwise.

\item
The rules dealing with predefined functions (namely, \textsc{predef,
  ifthentrue, ifthenfalse, andtrue, andfalse}) will be applied to the
expression returned by the previous step as many times as possible.
\end{enumerate}
 
\noindent
The proof will be driven by induction on the length of the trace for
$\mathit{goal}$.

{\bf Base case:} ($[(\Lambda \equiv f~p_1 \ldots
  p_a|g=r,o=\{\})] \in \mathit{Tr}(I_\infty,\mathit{goal})$). 

If $\mathit{goal}$ can be rewritten to normal form $e'$ by just using the rule
$\Lambda$, that means that $\Lambda \neq \Lambda_{\sbot,f}$ (since
$e'$ is assumed to be in normal form) and that $I_\infty$ must
countain a fact $f~t_1 \ldots t_a|c=b$, acording to the definition for
$\mathit{tr'}$ (second case), such that $\exists
\sigma'=\mathit{mgu}((t_1,\ldots,t_a),(e^+_1,\ldots,e^+_a))$ and
$\mathit{eval}(I_\infty,\sigma'(c))=\mathit{True}$ and
$\mathit{eval}(I_\infty,\sigma'(b))=e', ~~~~
(e^+_i=\mathit{eval}(I_\infty,e_i))$.

On the other hand, $\leadsto$ can apply \textsc{rule} to
$\mathit{goal}=f~e_1 \ldots e_a$. This rule rewrites $\mathit{goal}$
to: 

\[
  \sigma(g \MyIfThenOp r)
\]

\noindent
  where $\sigma=\mathit{mgu}((p_1,\ldots,p_a), (e^+_1,\ldots,e^+_a))$. Note that
  the application of $\mathit{eval}$ here is equivalent to the
  application of all the cases of $\mathit{uevalAux}$ except the last two ones 
(let us call these first cases $\mathit{uevalAux_\mathit{predef}}$) which in turn have the same effect
than the $\leadsto$-rules \textsc{predef, ifthentrue, ifthenfalse, andtrue,
    andfalse}  
as many times as necessary to evaluate any predefined
  function that appears in any of the arguments to $f$.
Let us remark that $\mathit{match}$ is unnecessary in $\leadsto$
since all the expressions handled by $\leadsto$ are ground and the substitution
returned by $\mathit{match}$ generates the same ground expressions than 
$\sigma$ in rule \textsc{rule}. We also assume that $\mathit{match}$ cannot be used in normal programs.

The same cases for predefined functions ($\mathit{uevalAux_\mathit{predef}}$) can be applied to the
expression above to get $\sigma(\mathit{eval}(I_\infty,g) \MyIfThenOp
\mathit{eval}(I_\infty,r))$.

By construction, we know that
$\sigma'(t_1,\ldots,t_a)=\sigma(p_1,\ldots,p_a)=(e^+_1,\ldots,e^+_a)$. Since
any valid program in our setting can only have at most one rule that
matches the ground expression $f e_1 \ldots e_n$, then:

\begin{itemize}
\item
$\mathit{eval}(I_\infty,\sigma'(b))=\mathit{eval}(I_\infty,\sigma(r))=e'$

\item
$\mathit{eval}(I_\infty,\sigma'(c))=\mathit{eval}(I_\infty,\sigma(g))=\mathit{True}$
\end{itemize}

\noindent
Therefore, both unfolding and $\leadsto$ have used a fact and a rule
which were sintactically indentical (once predefined functions have
been evaluated) to find the same answer for $\mathit{goal}$. This
proves the base case. 

{\bf Induction step:} Now the length of the trace for $\mathit{goal}$
is equal to $l+1 ~(l>0)$. Let us assume, as induction hypothesis that:

\[
  \begin{split}
  e' \in \mathit{ueval}(I_\infty,\mathit{goal}) \rightarrow
  \mathit{goal} \leadsto^* e' \\
  ~~ \text{provided that the trace for
    $\mathit{goal}$ has exactly $l$ elements} \\
    \text{and using the sequence of $\leadsto$-rules given earlier
      for that trace}) 
  \end{split}
\]

\noindent
Let us prove the equation for goals with trace of length $l+1$. In order
to do that let us consider that
$ [(\Lambda_1,o_1),\ldots,(\Lambda_{l+1},o_{l+1})] \in \mathit{Tr}(I_\infty,\mathit{goal})$
and an intermediate expression $e^l$ whose trace is the same as before
except for the first element.

Two cases have to be looked at here: One in which \textsc{rulebot} is
applied as first step (that is, $\Lambda_1=\Lambda_{\sbot,f'}$ for some
$f'$) and one more where \textsc{rule} is applied as first step (that
is, $\Lambda_1 \neq \Lambda_{\sbot,f'}$).

Let us begin with \textsc{rulebot}. If
$e^l=\mathit{goal}[\sbot]|_{o_1}$, the following must be true:

\begin{itemize}
\item
$\mathit{goal}|_{o_1}=f''~e''_1 \ldots e''_f ~~(f'' \in \FS_f)$

\item
$\nexists (F'' \in I_\infty~\text{such that}~F'' \equiv f''~t''_1
  \ldots
  t''_f|c''=b''~\text{and}$ \\
  $\sigma'=\mathit{mgu}((t''_1,\ldots,t''_f),(e''^+_1,\ldots,e''^+_f))
  ~ \text{ where
  }~e''^+_j=\mathit{eval}(I_\infty,e''_j)\\~\text{and}~\mathit{True} \in
  \mathit{ueval}(I_\infty,\sigma'(c'')))$. 

\item
If $F''$ does not exist that means that no rule for $f''$ originates a
fact like $F''$ which in turn means that no rule unifies with
$(e''^+_1,\ldots,e''^+_f)$ either. This forces $\leadsto$ to apply
\textsc{rulebot} to $e^l|_{o_1}$ and the same does $\mathit{ueval}$
(last rule).
\end{itemize} 

\noindent
Note that the sequence of $\leadsto$ rules applied in this case was
\textsc{(\textsc{predef, ifthentrue, ifthenfalse, andtrue,
    andfalse})$^*$, \textsc{rulebot}}.  

Since the trace for $\mathit{goal}[\sbot]|_{o_1}$ has length $l$, the
induction hypothesis is suitable for it and then:

\[
  \begin{split}
  e' \in \mathit{ueval}(I_\infty,e^l) \rightarrow e^l \leadsto^* e' \\
  \text{by using the sequence of $\leadsto$-rules given earlier for
  $e^l$})
  \end{split}
\]

\noindent
and since the normal form for $\mathit{goal}$ is the same as the
normal form for $e^l$ by definition of unfolding, we have that the
trace for $\mathit{goal}$ has the required shape.

Lastly, consider some goal whose trace is again \\
$[(\Lambda_1, o_1), \ldots, (\Lambda_{l+1}, o_{l+1})]$ but where
$\Lambda_1 \neq \Lambda_{\sbot,f}$. Let us apply $\mathit{uevalAux_\mathit{predef}}$
(or, equivalently,
\textsc{predef, ifthentrue, ifthenfalse, andtrue, andfalse}) as many times as
possible to the arguments of $\mathit{goal}|_{o_1}$ to get
$f''~e''^+_1 \ldots e''^+_f$ and then \textsc{rule} (the
only rule that $\leadsto$ can apply to $f''~e''^+_1
\ldots e''^+_f$) by using the rule $\Lambda_1 \equiv f''~p''_1 \ldots
p''_f|g''=r''$ and the unifier \\
$\sigma=\mathit{mgu}((p''_1,\ldots,p''_f), (e''^+_1,\ldots,e''^+_f))$. Now, we can rewrite $\mathit{goal}$ to: 

\[
  \begin{split}
  \mathit{goal}[\sigma(g'' \MyIfThenOp r'')]|_{o_1}
  \end{split}
\]

\noindent
given that according to $\mathit{tr'}$'s definition (second case), the  
rule $\Lambda_1$ is applicable to $\mathit{goal}|_{o_1}$ (since the
trace of any fact begins by the rule which originates the 
fact and $\mathit{unfold}$ has been able to apply a fact derived from
$\Lambda_1$ to $\mathit{goal}|_{o_1}$). This means that the expression
above  can be rewritten to  
$\mathit{goal}[\mathit{eval}(I_\infty,\sigma(g'' \MyIfThenOp r''))]|_{o_1}$
which is the new $e^l$. Note that the $\mathit{eval}$ in the former
expression is equivalent to (\textsc{predef, ifthentrue, ifthenfalse, andtrue,
    andfalse})$^*$ (which in turn is equivalent to applying the cases of $\mathit{uevalAux_\mathit{predef}}$
 as many times as necessary).

What we know now is:

\begin{itemize}
\item
The normal form for $\mathit{goal}$ is the same as the normal form for
$e^l$ since $\leadsto$ conserves the semantics by definition.

\item
The rewriting sequence from $\mathit{goal}$ to its normal form is the
same as the rewriting sequence for $e^l$ preceded by the rules applied
above (i.e. \textsc{(\textsc{predef, ifthentrue, ifthenfalse, andtrue,
    andfalse})$^*$, rule, (\textsc{predef, ifthentrue, ifthenfalse, andtrue,
    andfalse})$^*$}).

\item
Since the length of the trace for $e^l$ is equal to $l$, the induction
hypothesis is applieable to it and then:

\[
  \begin{split}
  e' \in \mathit{ueval}(I_\infty,e^l) \rightarrow e^l \leadsto^* e' \\
  \text{by using the sequence of $\leadsto$-rules given earlier for $e^l$})
  \end{split}
\]
\end{itemize}

\noindent
Since we have proved that, given an unfolding sequence, we are able to
find a precise sequence of $\leadsto$-rules that provokes the very
same effect to any given expression, we have also proved the
$\leftarrow)$ implication of the theorem.

\subsection{Example}

Let us present an example that may clarify some of the concepts
involved in the proof of equivalence. Consider this program:

\lstset{style=bluecode}
\begin{lstlisting}
f(x)=g(x+1)
g(x)=h(x+2)
h(x)=x+3

j(5)=6

f2(x,y)=x+y

goal=f(4)
goal2=f2(f(4),f(4))
goal3=K(j(5))
\end{lstlisting}
\lstset{style=code}

\noindent
What we have here are a constructor (\lstinline.K.), some test goals
(\lstinline.goal, goal2. and \lstinline.goal3.) and some functions. A
first function group (\lstinline.f,g,h.) is such that \lstinline.f. requires the
evaluation of \lstinline.g. and \lstinline.h.. We will see that the
trace for \lstinline.f.'s facts reflects this issue. Next, we have two
more unrelated functions. \lstinline.j. and \lstinline.goal3. will be
used to demonstrate the usage of \textsc{rulebot} while
\lstinline.f2. and \lstinline.goal2. will show why non-unique traces
exist.

The first interpretation generated by the unfolder is:

\lstset{style=bluecode}
\begin{lstlisting}
* h(b) = b+3  <H>
* j(5) = 6  <J>
* f2(b,c) = b+c  <F2>
* goal3 = K(Bot)  <Goal3>
\end{lstlisting}
\lstset{style=code}

\noindent
The sequences enclosed between \lstinline.<. and \lstinline.>. is the
trace of every fact. It can be seen, however, that the unfolder does
not display the position of every step and it does not display the
usage of $\Lambda_{\sbot,j}$ rules either (see the last fact).

So, the whole trace for the last fact would be \\
\lstinline.[(Goal3,{}), (Lambda_Bot,{1})].. 

All the other facts have a trace of length 1. It can be seen that they
are identical to their respective program rules. This property
supports the proof for the base case of the induction since applying
the rule to perform rewriting is exactly the same as applying a fact
with a trace length of one. Realize that even though these facts
cannot be unfolded any further, they require $\mathit{eval}$ to reach
a normal form (since the addition needs to be evaluated after its
arguments have been bound to ground values).

Further interpretations provide more interesting traces. This is
$I_4$:

\lstset{style=bluecode}
\begin{lstlisting}
* h(b) = b+3  <H>
* j(5) = 6  <J>
* f2(b,c) = b+c  <F2>
* g(b) = b+2+3  <G,H>
* goal3 = K(6)  <Goal3,J>
* f(b) = b+1+2+3  <F,G,H>
* goal = 10  <Goal,F,G,H>
* goal2 = 20  <Goal2,F2,F,G,H,F,G,H>
\end{lstlisting}
\lstset{style=code}

\noindent
Take a look at how the fact regarding \lstinline.goal. reflects the
dependency between \lstinline.f., \lstinline.g. and \lstinline.h.. That
fact has a trace of length 4 (it is very easy to follow how
\lstinline.goal. is evaluated by looking at its trace). Removing the
first element of its trace (as  needed in the induction step) yields
the trace \lstinline.<F,G,H>. for which there is a fact (the fact for
\lstinline.f.). This means that in this case, the induction step says
that evaluating \lstinline.goal. is the same as applying \textsc{rule}
to find \lstinline.f(4). ($\mathit{eval}$ is not needed in this 
step) and then applying the induction hypothesis to
\lstinline.f(4). whose trace is an element shorter than that of
\lstinline.goal.. 

Finally, consider how \lstinline.goal2. can be evaluated to normal
form in multiple orders. Since \lstinline.f2. demands both arguments,
both of them must be taken to normal form but the order in which is
done is irrelevant. Since our unfolder does not show the positions for
the reduction steps all the traces for \lstinline.goal2. look the same
but more than one trace would appear if positions were taken into account.

\subsection{Lemma: Applicability of Every Step Inside a Fact's Trace}

We have seen that the proof of equivalence between the operational
semantics and the unfolding semantics relies on the fact that every
element inside a fact's trace is applicable to the expression resulting
from applying all the trace steps that preceded the steps under
consideration. How can we be sure that a trace step is always
applicable to the expression on which that step must operate?. This
lemma states that this will always happen. Intuituvely, what the lemma
says is that the trace of a fact is nothing else that the sequence
of rules that $\mathit{unfold}$ has applied to get from a program rule
to a valid fact.

\begin{lemma}[Applicability of Every Step Inside a Fact's Trace]
Let $e$ be an expression and let
$[(\Lambda_1,o_1),\ldots,(\Lambda_m,o_m)] \in
\mathit{Tr}(I_\infty,e)$. If $e \leadsto^* e_{m'} ~~(m' \geq 0, m'<m)$
and $[(\Lambda_{m'+1},o_{m'+1}),\ldots,(\Lambda_m,o_m)] \in
\mathit{Tr}(I_\infty,e_{m'})$ then the following assertions are true:

\begin{enumerate}
\item
If $\Lambda_{m'+1}=\Lambda_{\sbot,f'}$ then $\nexists (\Lambda \in
\mathit{rule}(P)=f'~t'_1 \ldots t'_n|g'=r'$ such that $\exists
\sigma=\mathit{mgu}(f'~t'_1 \ldots t'_n,\mathit{ueval}(I_\infty,e_{m'}|_{o_{m'+1}}))$ and
$\mathit{True} \in \mathit{ueval}(I_\infty,\sigma(g')))$

\item
If $\Lambda_{m'+1} \neq \Lambda_{\sbot,f'}$ then $\Lambda_{m'+1} \equiv f~t_1
\ldots t_n|g=r$ and \\ 
$\exists \sigma=\mathit{mgu}(f~t_1 \ldots
t_n,\mathit{eval(I_\infty,e_{m'}|_{o_{m'+1}}}))$ and $\mathit{True} \in
\mathit{ueval}(I_\infty,\sigma(g))$
\end{enumerate}


\end{lemma}

\section{Additional Examples}
\label{sec_additional_examples}

\begin{example}[Lazy Evaluation]
\label{ones_example}

Think of the following code and its first interpretations:

\begin{tabular}{m{3.75cm} m{8.25cm}}
\lstset{style=bluecode}
\begin{lstlisting}
first : [a] -> a
first (x:_) = x

ones : [Int]
ones = 1:ones

main : Int
main = first ones
\end{lstlisting}
\lstset{style=code}

&

\begin{tabular}{m{0.5cm} m{0.15cm} m{6.5cm}} 
$I_0$ & $=$ & $I^\bot_0=\emptyset$
\\
$I_1$ & $=$ & $U(I_0)=\{\lstinline.ones=1:.\sbot,$ \\
 & & $\lstinline.first(x:xs)=x.\}$
\\
$I^\bot_1$ & $=$ & $\{\lstinline.main=.\sbot \}$
\\
$I_2$ & $=$ & $U(I_1)=\{\lstinline.ones=1:(1:.\sbot\lstinline.).,$
\\
& & \lstinline.first(x:xs)=x, main=1.$\}$ 
\\
$I^\bot_2$ & $=$ & $I^\bot_1$
\end{tabular} 
\\

\end{tabular}

\noindent
The semantics for this program is infinite: every step adds
a \lstinline.1. to the list generated by \lstinline.ones..

Consider the step from $I_1$ to $I_2$: when unfolding
\lstinline.ones., a fact matching \lstinline.ones. is found in $I_1$ (namely,
\lstinline.ones=1:.$\sbot$) so this last value is replaced in the right side of
the rule. Since the new value for \lstinline.ones. is {\em greater} than the
existing fact and both heads are a variant of each other, the function
{\em clean} can remove the old fact.The fact
\lstinline.ones=1:.$\sbot$ can now be used to 
evaluate \lstinline.main.. Since the new fact  
\lstinline.main=1. is greater than the fact
\lstinline.main=.$\sbot$ it replaces the existing one. The
fact for \lstinline.first. remains unaltered.

\end{example}

\begin{example}[Larger Example]
\label{ex_global}

Let us present an example that intends to describe how all the
functions and concepts that we have seen throughout the paper
work. Think of the following program:

\lstset{style=bluecode}
\begin{lstlisting}
ite : Bool * a *  a -> a
ite(True,t,e) = t
ite(False,t,e) = e

gen : Int -> [Int]
gen n = n:(gen (n+1))

senior : Int -> Bool
senior age = ite(age>64,True,False)

map : (a -> b) *  [a] -> [b]
map(f,[])=[]
map(f,(x:xs)) = (f x) : (map(f,xs))

main50 : [Bool]
main50 = map(senior,gen(64)) 
\end{lstlisting}
\lstset{style=code}

\noindent
Let us see how this program is unfolded.

First, the initial interpretation ($I_0$) is empty, by definition. At
this point, the function {\em unfold} is applied to every rule in
turn, using $I_0$ as the second argument. This produces the following
interpretation ($I_1$):

\lstset{style=bluecode}
\begin{lstlisting}
* ite(True,b,c) = b
* ite(False,b,c) = c
* gen(b) = Cons(b,Bot)
* map(b,Nil) = Nil
* map(b,Cons(c,d)) = Cons(b@[c],Bot)
\end{lstlisting}
\lstset{style=code}

\item
How did we get here?. When a rule is applied to {\em unfold}, every
full application of a symbol in FS is replaced by the value assigned
to the application in the interpretation also applied to {\em
unfold}. The actual matching between an expression and some rule
head is performed by {\em umatch}, which is called by {\em unfold}
every time an expression needs to be unfolded. In most cases, {\em
  umatch} behaves as a simple unifier calculator but higher order
brings complexity into this function (in \emph{s}-semantics, where
higher order does not exist, simple unification is used in the place of
{\em umatch}). In this case, the interpretation
applied was the empty one, so, the following has happened to every rule:

\begin{itemize}
\item
The two rules for \lstinline.ite. have no applications of user-defined
functions, so nothing has to be done to them in order to reach a fact
in normal form. That is why they appear in $I_1$ right away.

\item
The rule for \lstinline.gen. is a little bit different since this rule
does have an application to a user-defined function. However, since
$I_0$ contains nothing about those functions, all that {\em unfold}
can do is to replace that invocation by the special
symbol \lstinline.Bot. (represented by \sbot ~in formulas) to represent
that nothing is known about the value of \lstinline.gen (b+1)..

\item
The function \lstinline.senior. has no facts inside $I_1$ since the
function {\em clean} removes any unguarded fact with a body equal
to \sbot. This is precisely what has happened since $I_0$ contains no
information about \lstinline.ite., so the resulting new fact
for \lstinline.senior. would be \lstinline.* senior age=Bot.

\item
The first rule for \lstinline.map. is left untouched since it has no full
applications of user-defined functions (as it happened
with \lstinline.ite.).

\item
The second rule for \lstinline.map. generates the fact \\
\lstinline.* map(b,Cons(c,d)) = Cons(b@[c],Bot). where
the \lstinline.Bot. denotes that the value
for \lstinline.map(f,xs). is not contained inside $I_0$. 

\item
And, finally, there is not any fact for \lstinline.main50. since the
whole application of \lstinline.map. that appears at the root of the body
is unknown, so it gets replaced by \lstinline.Bot., which is in turn
eliminated by {\em clean} (and moved into $I^\bot_1$).
\end{itemize}

\noindent
Since we saw that two facts were removed by {\em clean} because they
did not have a guard and their body was equal to \lstinline.Bot.,
$I^\bot_1$, has the following content:

\lstset{style=bluecode}
\begin{lstlisting}
* senior(age) = Bot
* main50 = Bot
\end{lstlisting}
\lstset{style=code} 

\noindent
These two facts will be {\em reinjected} into the factset when $I_2$
is calculated but in this case, they do not have a noticeable effect
on the results, so we will not insist on them any more. 

One more iteration of the unfolding operator generates $I_2$:

\lstset{style=bluecode}
\begin{lstlisting}
* ite(True,b,c) = b
* ite(False,b,c) = c
* map(b,Nil) = Nil
* gen(b) = Cons(b,Cons(b+1,Bot))
* senior(b)  |  snd(match(True,b>64)) = True
* senior(b)  |  snd(match(False,b>64)) = False
* map(b,Cons(c,Nil)) = Cons(b@[c],Nil)
* map(b,Cons(c,Cons(d,e))) = 
    Cons(b@[c],Cons(b@[d],Bot))
* main50 = Cons(Bot,Bot)
\end{lstlisting}
\lstset{style=code}

\noindent
Remember that $I_2$ has been calculated by taking $I_1 \cup I^\bot_1$
as the relevant interpretation. By definition of the unfolding
operator, $I_2$ includes all the facts that were already present inside
$I_1$ (unless they are removed by {\em clean}). 

Remember also that we are using the optimized version of {\em clean}
(the one that removes subsumed facts instead of enlarging the  
constraints of the subsuming facts). Once these aspects have been
settled, the calculations that lead to the formation of $I_2$ 
can be explained as follows:

\begin{itemize}
\item
The two facts for \lstinline.ite. are transferred directly from $I_1$
into $I_2$. This is so since they cannot be unfolded any further and 
besides, they are not overlapped by any fact. The same happens with
the first fact for \lstinline.map.. 

\item
The fact for \lstinline.gen. is much more interesting: There are not
two facts for \lstinline.gen. in $I_2$. There is only one. This is due
to the application of {\em clean} in {\em unfold}. What has happened
here is that {\em clean} has compared the old fact 
(\lstinline.* gen(b)=Cons(b,Bot).) to the new one \\
(\lstinline.* gen(b)=Cons(b,Cons(b+1,Bot)).) and has removed the old
one. The reason for this is that both facts clearly 
overlap but the newest fact has a body that is greater (according to
$\sqsubseteq$) than that of the old fact. Given that the optimized 
version of {\em clean} is being used (all the functions here are
complete and the rules are productive), the old fact is removed. 

One more point of interest here: Note that the
expression \lstinline.b+1. cannot be further unfolded since the value
for \lstinline.b. is unknown at unfolding time. We will see the
opposite case later. 

\item
The explanation for \lstinline.senior. will be detailed later.

\item
The two facts for \lstinline.map. have become three. This has happened
as follows: 

\begin{itemize} 
\item
The second rule for \lstinline.map., when unfolded using $I_1 \cup
I^\bot_1$ generates two facts:  

\lstset{style=bluecode}
\begin{lstlisting}
* map(b,Cons(c,Nil)) = Cons(b@[c],Nil)
* map(b,Cons(c,Cons(d,e))) = Cons(b@[c],Cons(b@[d],Bot))
\end{lstlisting}
\lstset{style=code}

\noindent
Those two facts overlap with the old fact\\
 (\lstinline.* map(b,Cons(c,d)) = Cons(b@[c],Bot).), so this fact is
 removed by {\em clean}, which brings us to the count of three facts
 for \lstinline.map.. 
\end{itemize} 

\item
\lstinline.main50. has progressed slightly: The invocation
 of \lstinline.map. within the body of \lstinline.main50. has been
 replaced by the body of the second fact for \lstinline.map. in
 $I_1 \cup I^\bot_1$ generating  
\lstinline.Cons(senior@[b],Bot).. Since nothing is known about 
\lstinline.senior. in $I_1 \cup I^\bot_1$, the final result
 is \lstinline.Cons(Bot,Bot).. 
\end{itemize}

\noindent
The unfolding of \lstinline.senior. requires special attention: In
order to unfold the only rule for this function, the call 
to \lstinline.ite. is unfolded. However, the first argument
for \lstinline.ite. must be fully known before proceeding. This 
is impossible at unfolding time since \lstinline.age. will receive its
value later, at runtime. The only way to go in cases like 
this is to assume certain hypotheses and to generate facts that record
those hypotheses. In this example, we are forced 
to assume that \lstinline.age>64. is \lstinline.True. when the first
rule for \lstinline.ite. is unfolded while \lstinline.age>64. 
is assumed to be \lstinline.False. when the second rule
for \lstinline.ite. is unfolded. These hypotheses are recorded in 
the guards for the facts corresponding to \lstinline.senior..

The function responsible for generating these hypotheses is {\em
umatch} (more specifically, its second rule). This 
rule is used when an expression rooted by a predefined function
(here, \lstinline.<.) has to be matched to some pattern term  
which is not a variable (here, \lstinline.True. and
then \lstinline.False.). In this case, {\em umatch} extends the new
fact's guard by adding the new condition
(here \lstinline.snd(match(True,b>64)).) (resp. \lstinline.False.) and
then proceeds as if the PF-rooted expression matched the given pattern
in order to continue generating hypotheses. In this case, 
{\em umatch} would call itself with {\em umatch(True,True)},
(resp. {\em False}) which is solved by using {\em umatch}'s first rule
which generates no more conditions or variable substitutions.

Unfolding once again yields $I_3$:

\lstset{style=bluecode}
\begin{lstlisting}
* ite(True,b,c) = b  
* ite(False,b,c) = c  
* map(b,Nil) = Nil 
* senior(b) | snd(match(True,b>64)) = True  
* senior(b) | snd(match(False,b>64)) = False  
* map(b,Cons(c,Nil)) = Cons(b@[c],Nil) 
* gen(b) = Cons(b,Cons(b+1,Cons(b+1+1,Bot))) 
* map(b,Cons(c,Cons(d,Nil))) = Cons(b@[c],Cons(b@[d],Nil)) 
* map(b,Cons(c,Cons(d,Cons(e,f)))) = 
    Cons(b@[c],Cons(b@[d],Cons(b@[e],Bot))) 
* main50 | snd(match(True,64>64)),
           snd(match(True,65>64)) = Cons(True,Cons(True,Bot)) 
* main50 | snd(match(True,64>64)),snd(match(False,65>64)) = 
     Cons(True,Cons(False,Bot)) 
* main50 | snd(match(True,64>64)) = Cons(True,Cons(Bot,Bot)) 
* main50 | snd(match(False,64>64)),
      snd(match(True,65>64)) = Cons(False,Cons(True,Bot)) 
* main50 | snd(match(False,64>64)),
      snd(match(False,65>64)) = Cons(False,Cons(False,Bot)) 
* main50 | snd(match(False,64>64)) = Cons(False,Cons(Bot,Bot))
* main50 | snd(match(True,65>64)) = Cons(Bot,Cons(True,Bot)) 
* main50 | snd(match(False,65>64)) = Cons(Bot,Cons(False,Bot))
* main50 = Cons(Bot,Cons(Bot,Bot))  
* main50|snd(match(True,65>64)),snd(match(True,64>64))= 
     Cons(True,Cons(True,Bot))  
* main50 | snd(match(True,65>64)),
      snd(match(False,64>64)) = Cons(False,Cons(True,Bot)) 
* main50 | snd(match(False,65>64)),
     snd(match(True,64>64)) = Cons(True,Cons(False,Bot)) 
* main50 | snd(match(False,65>64)),
     snd(match(False,64>64)) = Cons(False,Cons(False,Bot)) 
\end{lstlisting}
\lstset{style=code}

\noindent
We are not repeating all the details above. Instead, we just want to
point out some interesting aspects of this interpretation: 

\begin{itemize}
\item
The reader might have expected to find expressions
like \lstinline.64>64. fully reduced (that is, replaced
by \lstinline.False.). That would be correct but boolean operators are
not evaluated due to a limitation in the implementation of our
unfolder. In this example, this limitation is a blessing in disguise
since those expressions are needed to understand the origin of some facts. 

\item
An expression like \lstinline.b+1+1. has not been reduced
to \lstinline.b+2. since it stands for \lstinline.(b+1)+1.. The
function {\em eval} has returned the same expression that it is given
since it cannot be further evaluated. 

\item
The combinatory explosion of facts denotes that the unfolder tries all
possible unfolding alternatives (in particular, those facts with less
than two conditions in the guard are the result of
unfolding \lstinline.senior. before \lstinline.ite., so the result for
\lstinline.senior. cannot be other than an unguarded $\sbot$).

\item
Note that our Prolog implementation does not have an underlying
constraint solver, so the entailment condition of the guards that is
used to sort overlapping facts is not checked. That is why the
unfolder has generated facts that should have been removed, such as
\lstinline.main50 = Cons(Bot ,Cons(Bot ,Bot ))..   

\item
A value of \lstinline.65. appears whenever the
function \lstinline.eval. has been applied to evaluate 64+1. 

\end{itemize}

\end{example}

\begin{example}[Unfolding and Abstract Interpretation]
\label{ex_abs1}
This example will show how unfolding can be used to synthesize an
abstract interpreter of a functional program. Think of the problem of
the parity of addition. The sum of Peano naturals can be defined as
shown in Fig.\ref{fig_original} (right). 

We also know that the successor of an even number is an odd number and
viceversa. The abstract domain (the domain of parities) can be
written as:

\lstset{style=bluecode}
\begin{lstlisting}
data Nat# = Suc_c# Nat# | Even# | Odd#
\end{lstlisting}
\lstset{style=code}

\noindent
Now, the user would define the abstract version for
\lstinline.add. together with the properties of
\lstinline.Suc. regarding parity:

\lstset{style=bluecode}
\begin{lstlisting}
add# : Nat# -> Nat# -> Nat#
add# Even# m = m 
add# (Suc_c# n) m = Suc_f# (add# n m)

Suc_f# : Nat# -> Nat#
Suc_f# Even# = Odd#
Suc_f# Odd# = Even#
\end{lstlisting}
\lstset{style=code}

\noindent
In order to enforce the properties of the successor in the abstract
domain, a catamorphism \footnote{A catamorphism takes a term an
  returns the term after replacing constructors by a corresponding
operator.} linking \lstinline.Suc_f#. to \lstinline.Suc_c#. will be used:

\lstset{style=bluecode}
\begin{lstlisting}
C_s : Nat# -> Nat#
C_s Even# = Even#
C_s Odd# = Odd#
C_s (Suc_c# n) = Suc_f# (C_s n)
\end{lstlisting}
\lstset{style=code}

\noindent
Then, the unfolding process that has been described must be slightly
modified: after every normal unfolding step, every abstract term in a
pattern must be replaced by the term returned by the
catamorphism. By doing this, the unfolding of the previous program
reaches a fixed point at $I_2$ \footnote{The rules for the
   catamorphism do not take part in unfolding}:

\lstset{style=bluecode}
\begin{lstlisting}
* add#(Even#,m) = m
* Suc_f#(Even#) = Odd#
* Suc_f#(Odd#) = Even#
* add#(Odd#,Odd#) = Even#
* add#(Odd#,Even#) = Odd#
\end{lstlisting}
\lstset{style=code}
\end{example}

\begin{example}[Addition of Parities Revisited]
\label{ex_parities_revisited}

As an interesting point of comparison, consider this alternative
version for \lstinline.add#.:

\lstset{style=bluecode}
\begin{lstlisting}
addr# : Nat# -> Nat# -> Nat#
addr# Even# m = m 
addr# (Suc_c# n) m = addr# n (Suc_c# m)
\end{lstlisting}
\lstset{style=code}

\noindent
The fixed point for this new function is as follows (also in $I_2$):

\lstset{style=bluecode}
\begin{lstlisting}
* addr#(Even#,b) = b
* addr#(Odd#,b) = Suc#(b)
* suc_f#(Even#) = Odd#
* suc_f#(Odd#) = Even#
\end{lstlisting}
\lstset{style=code}
\end{example}

\begin{example}[Demand Analysis]
\label{ex_demand_analysis}

The following example shows how abstraction can help to find program
properties. This particular example investigates how to find demand
properties for the functions in a program. By {\em demand properties}
we mean the level of definition that a function requires in its
arguments in order to return a result strictly more defined than
$\sbot$.

For the sake of simplicity, we are limiting our analysis to top-level
positions within the arguments although the method can be easily
extended to cope with deeper positions.

As before, we begin by defining the abstract domain. This example will
run on Peano Naturals, so the new domain reflects what elements are
free variables and what others are not:

\lstset{style=bluecode}
\begin{lstlisting}
data NatDemand# = Z# | S# NatDemand# | FreeNat# 
\end{lstlisting}
\lstset{style=code}

\noindent
As an example, we will use the well known function
\lstinline.leq.. \lstinline.leq x y. returns whether \lstinline.x. is
lesser or equal than \lstinline.y.. The standard (unabstracted)
version of \lstinline.leq. is as follows:

\lstset{style=bluecode}
\begin{lstlisting}
leq : Nat -> Nat -> Bool
leq Zero y = True
leq (Suc x) Zero = False
leq (Suc x) (Suc y) = leq x y
\end{lstlisting}
\lstset{style=code}

\noindent
The abstracted version, which is useful for finding the demand
properties for \lstinline.leq. at the top level positions of its
arguments is as follows:

\lstset{style=bluecode}
\begin{lstlisting}
data Bool# = True# | False# | DontCareBool#

leq# :: NatDemand# -> NatDemand# -> Bool#
leq# Zero# FreeNat# = DontCareBool#
leq# (Suc# x) Zero# = DontCareBool#
leq# (Suc# x) (Suc y) = leq# x y
\end{lstlisting}
\lstset{style=code}

\noindent
Observe that those rule bodies that do not influence the demand
properties of the function have been abstracted to
\lstinline.DontCareBool#. (and not to \lstinline.True#. and
\lstinline.False#. in order to get an abstract representation that is
as simple as possible while not losing any demand information). Note
that \lstinline.FreeNat#. represents that a certain argument is not
demanded. This abstraction transformation can be mechanised: Any
singleton variable in a rule is sure not to be demanded so it is
abstracted to \lstinline.FreeNat#.. The rest of variables are left as
they are.

What we need next is to define the functions that assert when a term
is not free (that is, demanded when it appears as a function
argument). We need one such function for every data constructor of
type \lstinline.NatDemand#.:

\lstset{style=bluecode}
\begin{lstlisting}
FreeNat_f# : NatDemand#
FreeNat_f# = FreeNat#

Z_f# : NatDemand#
Z_f# = Z#

S_f# : NatDemand# -> NatDemand#
S_f# FreeNat# = S# FreeNat#
S_f# Z# = S# Z#
S_f# (S# _) = S# FreeNat#
\end{lstlisting}
\lstset{style=code}

\noindent
We also need the catamorphsims that link the functions above to the
constructors belonging to the type \lstinline.NatDemand#.:

\lstset{style=bluecode}
\begin{lstlisting}
C_freeNat : NatDemand#
C_freeNat : FreeNat_f#

C_Z : NatDemand#
C_Z : Z_f#

C_S : NatDemand# -> NatDemand#
C_S (S# x) = S_f# (C_S x)
\end{lstlisting}
\lstset{style=code}

\noindent
As we did in the previous example, we now have to apply the following
steps to a program composed of the rules for \lstinline.leq#.,
\lstinline.freeNat_f#., \lstinline.Z_f#. and \lstinline.S_f#.:

\begin{itemize}
\item
Apply an unfolding iteration.

\item
Apply the catamorphisms to the heads of the resulting facts.

\item
Evaluate the resulting head expressions.
\end{itemize}

\noindent
The fixed point is reached at the second iteration ($I_2$). It contains
the following:

\lstset{style=bluecode}
\begin{lstlisting}
* leq#(Z#,FreeNat#) = DontCareBool#
* leq#(S#(FreeNat#),Z#) = DontCareBool#
* leq#(S#(FreeNat#),S#(FreeNat#)) = DontCareBool#

* z_f# = Z#

* s_f#(FreeNat#) = S#(FreeNat#)
* s_f#(Z#) = S#(FreeNat#)
* s_f#(S#(b)) = S#(FreeNat#)

* freeNat_f# = FreeNat#
\end{lstlisting}
\lstset{style=code}

\noindent
That means that \lstinline.leq#. does not demand its second argument
if the first one is \lstinline.Z#. (since
\lstinline.FreeNat#. represents no demand at all). However,
\lstinline.leq#. demands its second argument if the first one is
headed by \lstinline.S#.. Note that we are considering top level
positions for the arguments only but that deeper positions can be
easily considered by just extending \lstinline.s_f#..
\end{example}

\section{Validity of the Unfolded Code}

The lemma below supports the validity of the code generated by the
unfolding process:

\subsection{Proof of Lemma \ref{lemma_old_clean}}
Let $H$ be a fact generated by unfolding rule $\Lambda$
and belonging to interpretation $I_n$. Let $\{S_i~~(i:1..m)\}$ be the
set of facts that belong to $I_{n+1}$, that have been generated by
unfolding $\Lambda$ and which overlap with $H$.

By {\em reductio ad absurdum}, let us think that, even in the conditions
stated, the $S_i$  do not cover all the cases that $H$ covers. Then,
it must be possible to build at least one fact $S'$ that
overlaps with $H$ but that does not overlap with any fact $S_i$.

In order to build a fact like $S'$, the following options can be
taken:

\begin{enumerate}
\item
Choose $H$ such that its pattern and/or guard does not match with any
of the rules for $f$.

\item 
When unfolding $H$, use a fact that has not been used when calculating
the facts $\{S_i~~(i:1..m)\}$.
\end{enumerate}

\noindent
However, condition 1 is impossible since all the function definitions
are assumed to be complete (i.e. there is no fact for $f$ which does not match a
rule) and to have only generating rules. 
In addition, condition 2 is also impossible since {\em unfold}
 uses all the existing facts by definition.

Note that the condition which requires that the rules be generative
cannot be dropped since a complete function having one or
more non-generative rules would have some facts removed
from $I_{n+1}$ by {\em clean}, which would render 
the function definition incomplete in that interpretation.

Therefore, no fact like $S'$ can exist. We have reached a
contradiction and thus we have proved that
under the conditions stated for $P$, {\em clean} can always get rid of
the most general fact.

\subsection{Proof of Lemma \ref{lemma_clean_overlappings}}

If program $P$ does not have overlapping rules then any pair of rules
$l ~|~ g = r$ and $l' ~|~ g' = r'$ must meet
one of the following conditions:

\begin{enumerate}
\item
There is no unifier between $l$ and $l'$.

\item
If a substitution $\sigma$ is such that $\sigma=\mathit{mgu}(l,l')$,
then the constraint $g \wedge \sigma(g')$ is unsatisfiable.
\end{enumerate}

\noindent
At every application, the {\em unfold} function takes a
rule and applies a substitution to its pattern as well as a (possible)
conjunction to its guard. Now:

\begin{enumerate}
\item
If the two rules given do not overlap because $l$ and $l'$ cannot be
unified, applying any substitution to them makes them even less
unifiable.

\item
If the two rules given do not overlap because $l$ and $l'$ can be
unified but the conjunction of their guards cannot be satisfied, adding
a conjunction to either guards makes their combined satisfiability
even less likely.
\end{enumerate}

\noindent
Up to this point, we have shown that the unfoldings of any two non
overlapping rules cannot give rise to overlapping facts but the facts
generated by the unfolding of a single rule may still contain
overlapping pairs. In order to prove that the unfoldings of a single
rule from a program $P$ can be written without overlappings, we need
to use the function {\em ueval} that was defined in
Sect. \ref{sec_ueval}.

We now want to prove that, for any single rule $R$ belonging to a
program $P$ without overlapping rules, the unfoldings of $R$ carry the
same meaning with or without the cleaning phase. That is, let us call $P_R=\mathit{unfold}(R,I_\infty)$:

\begin{equation}
\label{eq_clean_rule}
  \mathit{ueval}(P_R,c \MyIfThenOp
  e)=\mathit{ueval}(\mathit{clean}(P_R),c \MyIfThenOp e)
  ~~~ \forall c \MyIfThenOp e \in E  
\end{equation}

\noindent
We will prove that Equation \ref{eq_clean_rule} holds by induction on
the number of full applications of symbols of $\FS$ held in $c$ and
$e$ combined.

{\bf Base case:} If neither $c$ nor $e$ have any full application of
symbols of $\FS$, then both $c$ and $e$ are expressions (terms which may include
calls to predefined functions) and therefore cannot
be unfolded any more. Their value (as computed by $\mathit{ueval}$)
does not depend on the interpretation used, so Equation
\ref{eq_clean_rule} trivially holds.

{\bf Induction step:} Let us assume that Equation \ref{eq_clean_rule}
holds if $c$ and $e$ have a combined total of $n$ full applications of
symbols of $\FS$ and let us try to prove that Equation
\ref{eq_clean_rule} holds when $c$ and $e$ have a combined total of
$n+1$ full applications of symbols of $\FS$.

In order to do that, let us define an expression $e'$ which has
exactly one more application of symbols of $\FS$ than $e$ (the
reasoning over $c$ would be analogous). Let us define $e'=e[f~t_1
  \ldots t_n]|_o$ where $e|_o \in E$ which no full invocations of
  symbols of $\FS$, $f \in \FS, t_i \in T$. This
guarantees that $e'$ has one more full application of symbols of $\FS$
than $e$. Since the induction hypothesis holds for $c \MyIfThenOp e$, all we
have to prove is:

\begin{equation}
\label{eq_f_lemma2}
  \begin{split}
  \mathit{ueval}(P_R,\text{True} \MyIfThenOp f~t_1 \ldots
  t_n)=
  \mathit{ueval}(\mathit{clean}_P(P_R),\text{True} \MyIfThenOp f~t_1 \ldots
  t_n)
  \end{split}
\end{equation}

\noindent
Now, if $P_R$ does not contain overlapping facts or does not contain
facts about $f$ at all, the Equation above trivially holds since the
interpretations $P_R$ and $\mathit{clean}_P(P_R)$ are the same by
definition of {\em clean}.

Let us now assume that $P_R$ contains (maybe among others), the
following facts:

\begin{itemize}
\item
$F \equiv f~p_1 \ldots p_n~|~c=b$

\item
$F_i \equiv \sigma_i(f~p_1 \ldots p_n~|~c \wedge c_i=b) ~~(i:1..m)$
\end{itemize} 

\noindent
That is, the facts $F_i$ overlap $F$ and $F_i$ are more specific than
$F$. Then, by definition of {\em clean}, $\mathit{clean}_P(P_R)$ will
hold the facts $F_i$ together with a new fact:

\[
\begin{split}
  F' \equiv f~p_1 \ldots p_n~|~c \wedge 
  \bigwedge_i(\mathit{nunif}((p_1,\ldots,p_n),\sigma_i(p_1,\ldots,p_n))
  \vee \mathit{not}(\sigma_i(c_i)))=b
\end{split}
\]

\noindent
Let $\chi=f~t_1 \ldots t_n$. The following cases can occur:

\begin{itemize}
\item
If $\chi$ is not unfoldable by $F$, then it is not unfoldable by any of
the more specific facts (the $F_i$ and $F'$), so Equation
\ref{eq_f_lemma2} holds.

\item
If $\chi$ is unfoldable by $F$ but not by any of the $F_i$, then $\chi$
is unfoldable by $F'$, which returns the same result as $F$.

\item
Lastly, if $\chi$ is unfoldable by $F$ and one of the $F_i$, then the
left side of Equation \ref{eq_f_lemma2} returns two values (let them
be $c^F \MyIfThenOp e^F$ and $c^F_i \MyIfThenOp e^F_i$) which verify
$c^F \MyIfThenOp e^F \sqsubseteq  c^F_i \MyIfThenOp e^F_i$. Since all
the functions have to be well-defined, the value for $\chi$ has to be
the greatest of the two mandatorily. The right side of
Equation \ref{eq_f_lemma2} returns only the value 
$c^F_i \MyIfThenOp e^F_i$  by definition of {\em clean} (which will have removed
$F$ from $P_R$ and replaced it by $F'$ which will not be usable to
unfold $\chi$).
\end{itemize}

\end{document}